\documentclass[12pt]{article}
\usepackage{amsfonts,amssymb,amsmath,amscd}
\usepackage{amsmath,amssymb,epsfig,ulem}
\newlength{\wth}
\newcommand{\startappendix}{
\setcounter{section}{0}
\renewcommand{\thesection}{\Alph{section}}}

\newcommand{\PP}{{\mathbb P}}
\newcommand{\ZZ}{{\mathbb Z}}

\newcommand{\RR}{{\mathbb R}}
\newcommand{\CC}{{\mathbb C}}

\newcommand{\VV}{{\mathbb V}}

\newcommand{\tf}{{\tilde f}}

\newcommand{\cO}{{\mathcal O}}

\newcommand{\tV}{{\widetilde V}}

\newcommand{\cV}{{\mathcal V}}

\newcommand{\cM}{{\mathcal M}}

\newcommand{\cN}{{\mathcal N}}

\newcommand{\cW}{{\mathcal W}}

\newcommand{\HH}{H_{\overline{D}, L^2}}

\newcommand{\tx}{{\tilde x}}

\newcommand{\bfV}{{\mathbf V}}
\def\be{\begin{equation}}
\def\ee{\end{equation}}
\def\bear{\begin{eqnarray}}
\def\eear{\end{eqnarray}}

\newcommand{\dL}{{\widehat\Lambda}}
\def\half{{ \frac{1}{2} }}

\def\dg{{\dagger}}

\def\a{{\alpha}}

\def\dg{{\dagger}}

\def\tC{{{\tilde C}}}

\def\tb{{{\tilde b}}}

\def\delbar{{\overline {\partial}}}

\def\vert{{|}}

\def\ts{{\tilde s}}
\def\tx{{\tilde x}}

\newcommand{\LG}{{{}^LG}}

\renewcommand{\thefootnote}{\fnsymbol{footnote}}
\begin{document}

\thispagestyle{empty}
\begin{flushright}\footnotesize
\texttt{pi-qf \& strings-287}\\
\vspace{2.1cm}
\end{flushright}

\renewcommand{\thefootnote}{\fnsymbol{footnote}}
\setcounter{footnote}{0}

\vspace{1.5cm}

\begin{center}{\bf \Large OPE of Wilson-'t Hooft operators in $\cN=4$ and $\cN=2$ SYM with gauge group $G=PSU(3)$}

\vspace{2.1cm}

\textrm{Ruxandra Moraru$^1$, Natalia Saulina$^2$}

\vspace{1cm}

\textit{$^1$ Pure Mathematics, University of Waterloo, ON, N2L 3G1, 
Canada}\\
\texttt{moraru, math.uwaterloo.ca}
\bigskip

\vspace{1cm}

\textit{$^2$Perimeter Institute, Waterloo, ON 
N2L2Y5, Canada}\\
\texttt{nsaulina, perimeterinstitute.ca}

\vspace{1cm}


\par\vspace{1cm}

\textbf{Abstract}\vspace{5mm}
\end{center}

\noindent
We compute the simplest non-trivial Operator Product Expansion of Wilson-'t Hooft loop operators
in $\cN=4$ and $\cN=2$ Super-Yang-Mills theory with gauge group $G=PSU(3)$.
This amounts to finding the Euler characters of certain  vector bundles, describing
electric degrees of freedom of loop operators entering the OPE,
over moduli spaces of BPS states in the presence of loop operators.

\vspace*{\fill}

\setcounter{page}{1}
\renewcommand{\thefootnote}{\arabic{footnote}}
\setcounter{footnote}{0}

 \newpage

\tableofcontents

\newpage

\section{Introduction}
Wilson loop operators \cite{Wilson} and 't Hooft loop operators \cite{GNO}, \cite{KWH} 
are famous examples of non-local observables in gauge theory. The Operator Product
Expansion (OPE) of these operators contains important information about the theory.
The product of parallel Wilson loops is determined by the
representation ring of the gauge group $G,$ while S-duality conjecture  \cite{MO}
predicts that product of parallel 't Hooft loops is controlled by the
representation ring of the Langlands dual group $\LG.$ This  prediction has been verified
in \cite{KW} based on the earlier mathematical result \cite{Lusztig}.

Yang-Mills theory also admits mixed Wilson-'t Hooft (WH) loop operators.
As explained in \cite{KWH}, at zero $\theta-$angle  they are labeled by elements of the set
$$
\dL(G)/\cW=\Bigl(\Lambda_w(G)\oplus \Lambda_w(\LG)\Bigr)/\cW,
$$
where $\Lambda_w(G)$ is the weight lattice of $G$ and $\cW$ is the
Weyl group (which is the same for $G$ and $\LG$). In $\cN=4$ or $\cN=2$ 
Super-Yang-Mills (SYM)
theory these mixed operators can be made supersymmetric preserving
 one quarter of the original supersymmetry.

In \cite{KS} an approach how to compute the product
of WH loop operators  in $\cN=2$ gauge theory
was outlined and the OPEs were
actually computed in $\cN=4$ SYM with gauge group $G=SU(2)$ and $G=PSU(2).$ 
More recently we determined the basic geometric ingredients \cite{Saulina} required for
the computation of the OPE of Wilson-t Hooft loop operators
 in $\cN=4$ SYM and $\cN=2$ SYM with gauge group $G=PSU(3).$
In this paper we use these ingredients to obtain the simplest non-trivial OPE of WH operators in $\cN=4$ and $\cN=2$ SYM theory for  $G=PSU(3)$:
 \be \label{goal}WT_{w_1,\, \nu}\times
WT_{w_1,\, 0}= WT_{2w_1, \, \nu}+\sum_j (-)^{s_j}\, WT_{w_2, \,\nu_j}\ee
where magnetic charge $w_1$($w_2$) is the highest weight of
a fundamental (anti-fundamental) representation
of $\LG=SU(3)$ and electric charge
$\nu=aw_1+bw_2$ is the highest weight of $G,$ i.e. $a+2b=0$ mod $3.$
The  electric weights $\nu_j$ and signs $(-)^{s_j}$ on the right side of (\ref{goal}) are 
explicitly determined in Section 3 for some values of $a,b$ and the prescription
how to compute them for general $a,b$ is provided.

Our approach uses the holomorphic-topological twist  \cite{htft} of
the $\cN=2$ gauge theory and the connection between BPS configurations 
in the presence of 't Hooft operators and solutions of 3d Bogomolny equations 
with magnetic sources \cite{KW},\cite{Witten2009}.
 To determine the right side of (\ref{goal}) in $\cN=4$ SYM,
  we have to
compute the Euler characters ${\bf I}_{\cN=4}\Bigl(\cM,\cV_{a,b}\Bigr)$ and ${\bf I}_{\cN=4}\Bigl(X,\cV^{bulk}_{a,b}\Bigr)$ of certain vector bundles $\cV_{a,b}$ and $\cV^{bulk}_{a,b}$
 on moduli spaces
$\cM$ and $X$ whose geometry was found in \cite{Saulina} and
is reviewed in Section 2. Similarly, to compute (\ref{goal}) in $\cN=2$ SYM we
compute the holomorphic Euler characters ${\bf I}_{\cN=2}\Bigl(\cM,\cV_{a,b}\Bigr)$ and ${\bf I}_{\cN=2}\Bigl(X,\cV^{bulk}_{a,b}\Bigr).$
The (holomorphic)  Euler characters compute with sign the
ground states of appropriate supersymmetric quantum mechanics (SQM)
with the BRST operator acting as the covariant Dolbeault  operator.
 This SQM arises as 
the result of quantizing  the gauge theory on $\RR\times I\times \mathcal{C}$ in the presence of Wilson-t' Hooft operators along $\RR.$ Here $I$ is an interval and $\mathcal{C}$
is a Riemann surface, and boundary conditions\footnote{For an explicit choice of
boundary conditions see \cite{KS}.} at the two
ends of $I$ are such that without any magnetic sources there is unique vacuum.
The WH operators $WT_{w_1,\, \nu}$ and $WT_{w_1,\, 0}$ are taken
to sit at the same point at $\mathcal{C}$ and at different\footnote{The twist \cite{htft}
ensures that there is no dependence on the distance between points along $I$} points along $I.$

The compact moduli space $\mathcal{M}$ is obtained by blowing-up
certain singular complex 4-fold  which is the compactification of the moduli space of solutions of 3d Bogomolny equations in $I\times \mathcal{C}$ with a single source
characterized by magnetic charge $2w_1.$ The blow-up procedure produces exceptional divisor $\mathcal{D}$ in $\mathcal{M}.$
The non-compact moduli space $X$ - referred to as
the `bulk part'  of $\cM$ - is obtained by removing
from $\mathcal{M}$ the vicinity of $\mathcal{D}$, i.e. the total space of the normal bundle of $\mathcal{D}$ in  $\mathcal{M}.$  As we review in Section 2, $\cM$ is a $\PP^2$ fibration over $\PP^2$ and $X=T\PP^2.$ In computing the Euler characters of bundles over $\cM$
(in Appendix A) we use  the Leray spectral sequence. Meanwhile, to compute these characters for bundles over $X$ we use $L^2$ Dolbeault cohomology for the metric 
written down in \cite{Saulina} and reviewed in equation (4) below.

The unknown electric weights $\nu_i$ and signs $s_i$
in the right side of (\ref{goal}) can be read off
from the so called bubbled contribution:
 $${\bf I}^{bubble}_{\cN=2}\Bigl(\cV_{a,b}\Bigr)={\bf I}_{\cN=2}\Bigl(\cM,\cV_{a,b}\Bigr)-
{\bf I}_{\cN=2}\Bigl(X,\cV^{bulk}_{a,b}\Bigr)$$
and
$${\bf I}^{bubble}_{\cN=4}\Bigl(\cV_{a,b}\Bigr)={\bf I}_{\cN=4}\Bigl(\cM,\cV_{a,b}\Bigr)-
{\bf I}_{\cN=4}\Bigl(X,\cV^{bulk}_{a,b}\Bigr)$$
decomposed into representations of the $SU(2)_{\alpha_1}\times U(1)$ which is a subgroup of the gauge group unbroken in the presence
of 't Hooft operator with magnetic charge $w_2.$
The existence of bubbled contribution is due to monopole
 bubbling \cite{KW} which occurs when the magnetic charge of the 't Hooft
 operator  decreases by absorbing a BPS monopole. This process is
 possible because the moduli space of solutions of 3d Bogomolny equations in the presence of magnetic source with charge $2w_1$ is non-compact.

Note that the computations in this paper can be viewed as a UV method.
There are alternative IR methods of studying loop operators \cite{GMN},\cite{ACCERV},\cite{Sen},\cite{GMN2}. Furthermore, loop operators
in a certain class \cite{Gaiotto} of $\cN=2$ SYM theories were studied
\cite{AGGTV},\cite{DGOT},\cite{DGOT2},\cite{Gomis},\cite{T} using connection with 2d Conformal Field Theory \cite{AGT}.  In the future 
we hope to compare our answers for the OPE of Wilson-'t Hooft operators 
in $\cN=2$ and $\cN=4$ SYM with gauge group $G=PSU(3)$ with the OPE which can be obtained from these alternative methods. 

This note is organized as follows. In Section 2 we review 
the geometry of  moduli spaces $\mathcal{M}$ and $X$ together with
vector bundles over them.  In Section 3 we explain how to compute the OPE (\ref{goal})
and provide explicit examples.
We compute the Euler characters for vector bundles on $\cM$ in Appendix A and,
under certain {\it vanishing assumption},
for vector bundles on $X$ in Appendix B.  We provide evidence supporting
the vanishing assumption in Appendix C and collect useful formulae
for computing the $L^2$ norms of differential forms taking values in line bundles on $X$ in Appendix D.

\section{Review of moduli spaces $\cM$ and $X$}
Here we review basic geometric ingredients \cite{Saulina} required for
the computation of the OPE of Wilson-t Hooft loop operators
 in $\cN=4$ SYM and $\cN=2$ SYM with gauge group $G=PSU(3).$
These include the geometry of $\mathcal{M}$ -  the moduli space of BPS configurations in the presence 
 of two 't Hooft operators each with fundamental magnetic weight, as well as 
 the geometry of its 'bulk part' - open subspace $X$ obtained by excision of the vicinity of the certain blown-up region in $\cM.$ We also present
vector bundles over $\cM$ and $X$ which encode electric degrees of freedom
of the loop operators in the OPE.

  It was shown in \cite{Saulina} that $\cM$ is defined by a hypersurface $y_aU^a=0$ in a  toric 5-fold $Y_5.$ The weights under the two $\mathbb{C}^*$ actions are
\be \label{action}\begin{array}{c|c|c|c|c|c|c|c}
&U^1 & U^2 & U^3 & \Lambda & y_1 & y_2 & y_3\cr
\hline
\text{first}&1 & 1 & 1& -2& 0& 0& 0 \cr
\hline
\text{second}&0& 0& 0& 1& 1&1&1\cr
\end{array}
\ee

The 5-fold $Y_5$  is a $\mathbb{P}^3$ fibration over
$\mathbb{P}^2.$ Here $U^1,U^2,U^3$ are homogenous coordinates on the base $\mathbb{P}_{\vec U}^2$
and $\Lambda, y_1,y_2,y_3$  are homogenous coordinates on the
fiber $\mathbb{P}^3.$ Meanwhile, the 4-fold $\cM$ 
is $\mathbb{P}^2$ fibration over $\mathbb{P}^2_{\vec U}.$ 
General $PSU(3)$ invariant metric on $\cM$ was written out  and confirmed
in \cite{Saulina} by reproducing correctly the OPE of 't Hooft operators.

Let $\pi: \cM\mapsto \PP^2_{\vec U}.$ The vector bundle $\cV_{a,b}$ on $\cM,$
which arises in computing  $WT_{w_1,\,\, aw_1+bw_2}WT_{w_1,0},$
is a pull-back from the base:
$$\cV_{a,b}=\pi^*\bfV_{a,b}$$
where $\bfV_{a,b}$ was determined in \cite{Saulina} to be
$$\bfV_{a,b}=\cO_{\PP^2}\Bigl(-(2a+b)\Bigr)\otimes S^{\vert b\vert}\tV_{1}$$
with $\tV_{1}$ a vector bundle on satisfying the properties
\be \label{info} r(\tV_1)=2,\quad c_1(\tV_1)=0,\quad c_2(\tV_1)=1,\quad H^0(\PP^2,\tV_1)=1,
\quad H^p(\PP^2,\tV_1)=0 \quad p=1,2.
\ee
These properties are a consequence of the fact that $\tV_1$ admits the following explicit connection in the patch $U^1\ne 0$ (see \cite{Saulina}, section 6):
 $$A_{(1,0)}=\Bigl(A_{(0,1)}\Bigr)^{\dg},\quad
A_{(0,1)}=i\left( \delbar \mathcal{G}\right)\mathcal{G}^{-1}\quad \quad
\mathcal{G}=\begin{pmatrix}\alpha &
\alpha\cr \alpha^{-1}\beta & \alpha^{-1}(1+\beta)\end{pmatrix}$$
with\footnote{Both choices of $\beta$ give the same connection. }
$$\alpha={y^{1/4}\over (y-1)^{1/2}},\quad \beta=-{{\bar z}_1\over (y-1)z_2} \quad \text{or} \quad \beta={{\bar z}_2\over (y-1)z_1},$$
where
$$y=1+\vert z_1\vert^2+\vert z_2\vert^2,\quad z_1={U^2\over U^1},\quad z_2={U^3\over U^1}.$$

Note that the bulk part $X$ of $\cM$ is $T\PP_{\vec U}^2.$ Indeed, $X$  was
defined in \cite{Saulina}, section 4, by $t_iU^i=0$ in the region $\Lambda\ne 0$ of $Y_5,$ where
$t_i={y_i\over \Lambda}$ is a (global) coordinate on the fibre of $\cO_{\PP^2}(2)$ for
each $i=1,2,3.$ 

A general $PSU(3)$ invariant K\"ahler form
 on $X$ was obtained in \cite{Saulina}, section 4:
 \be \label{k_form} (-iJ)=\sum_{J=1}^4 \tf_J(\ts) e_J\wedge \bar e_J\ee
 where in a patch $U^1\ne 0$
 $$\ts ={\tx\over y^2},\quad \tx=t_2 {\overline \alpha}^2+ t_3 {\overline \alpha}^3,\quad
y=1+\vert z_1\vert^2+\vert z_2\vert^2,\quad z_1={U^2\over U^1},\, z_2={U^3\over U^1}$$
 with
 $${\overline \alpha}^2=\bar t^2 +z^1 
 (\bar t^2 \bar z_1+\bar t^3 \bar z_2),\quad {\overline \alpha}^3=\bar t^3 +z^2 
 (\bar t^2 \bar z_1+\bar t^3 \bar z_2).$$
Explicitly, we may take\footnote{We use 
$e_1=-\ts^2 \mathcal{E}_1,\quad e_{2,4}=\sqrt{\ts}\mathcal{E}_{2,4},\quad
 e_3=\ts \mathcal{E}_3$ to relate $\tf_K(\ts)$ for $\ts=s^{-1}$ with $g_K(s)$  given
 in equation (27) in \cite{Saulina}.}
$$\tilde{f}_1(\ts)={1\over 2}{1\over \ts (1+\ts)^{3/2}},\quad
 \tilde{f}_2(\ts)={1\over \ts \sqrt{1+\ts}}+{1\over \ts},\quad
 \tilde{f}_3(\ts)={1\over \ts^2}(1-{1\over \sqrt{1+\ts}}),
 \quad \tilde{f}_4(\ts)={2\over \ts \sqrt{1+\ts}}. $$
This metric is used in Appendix C for explicit computation
of $L^2$ Dolbeault cohomology of vector bundles over $X.$

Note that the bundle $\cV^{bulk}_{a,b}$ which describes electric
degrees of freedom in Wilson-'t Hooft operator $WT_{2w_1, \,\, aw_1+bw_2}$
is again a pull back from the base $\PP^2_{\vec U}.$
We use the following  connection on $\cV^{bulk}_{-a,-b}$:
\be \label{def_con}\mathbb{A}_{(1,0)}=\Bigl(\mathbb{A}_{(0,1)}\Bigr)^{\dg},\quad
\mathbb{A}_{(0,1)}=i\left( \delbar
\mathcal{G}_{X}\right)\mathcal{G}_{X}^{-1}\quad
\quad \mathcal{G}_{X}={h^{(n)}(\ts)\over y^{n\over 2}}\,\,\mathcal{G}\quad \quad n=2a+b\ee
where the factor $h^{(n)}(\ts)$ in $\mathcal{G}_X$ describes the lift of the connection on 
$\cO_{\PP^2}(n)$ to $X$:
$$h^{(n)}(\ts)\sim \ts^{-n/4} \quad \ts\mapsto \infty,\quad h^{(n)}(\ts)\sim 1 \quad \ts\mapsto 0.$$
The
asymptotic at $\ts \rightarrow \infty$ is chosen in such a way
that the norm (evaluated at the point $z_1=z_2=0$
on the base $\PP_{\vec U}^2$) of the holomorphic section $s_{hol}=t_i$ of $\cO_X(2)$  approaches a
constant, i.e. we go to the unitary trivialization \be
s_{unit}=\mathcal{G}_X s_{hol}\label{transf} \ee and
require the pointwise norm $\vert s_{unit} \vert^2$ at $z_1=z_2=0$ to approach a
constant. The reason is that $\ts \mapsto  \infty$ limit corresponds
to approaching\footnote{$\ts \mapsto \infty$ near $\Lambda=0$ region in $\cM$} the blown-up region in $\cM$ and, as shown in \cite{Saulina}, $X$  behaves as
$\CC^2/\ZZ_2$ singularity fibered over $\PP_{\vec y}^2,$ i.e.
$\PP_{\vec U}^2$
effectively collapses to a point in this limit and, therefore, $t_i$ should become
a section of a trivial bundle.
 
The postulated behavior at $\ts \rightarrow 0$ (i.e. away from the blown-up region) ensures that the
connection on $X$ is the same as the connection on the total space $\cM$ -
a pull-back connection from $\PP^2_{\vec U}.$

\section{Computing OPE $WT_{w_1,aw_1+bw_2}WT_{w_1,0}$}
In the approach \cite{KS,Saulina}, to find the OPE $WT_{w_1,\,\, aw_1+bw_2}WT_{w_1,0}$ in $\cN=4$ SYM
we first have to compute the Euler characters of vector bundles on $\cM$ and $X$
$$\mathbf{I}_{\cN=4}(\cM,\cV_{a,b})= \sum_{\alpha=0}^4 (-)^{\a} I^{(\a)}\Bigl(\cM,\cV_{(a,b)}\Bigr),\quad
\mathbf{I}_{\cN=4}(X,\cV^{bulk}_{a,b})= \sum_{\alpha=0}^4 (-)^{\a} I^{(\a)}\Bigl(X,\cV^{bulk}_{(a,b)}\Bigr) $$
where for a vector bundle $V$ on a complex 4-fold $Y$ we denote
 $$I^{(\a)}(Y,V)=\sum_{p=0}^4 (-)^p \HH^p\Bigl(Y,\Omega^{\a}_Y \otimes V\Bigr).$$
 Note that Kodaira-Serre duality implies
 $$\mathbf{I}_{\cN=4}(\cM,\cV_{-a,-b})=\mathbf{I}_{\cN=4}(\cM,\cV_{a,b}),\quad
 \mathbf{I}_{\cN=4}(X,\cV^{bulk}_{-a,-b})=\mathbf{I}_{\cN=4}(X,\cV^{bulk}_{a,b}).$$
 
To determine the same OPE in $\cN=2$ SYM, we have to
compute the holomorphic Euler characters of vector bundles on $\cM$ and $X$
 $$I_{\cN=2}(\cM,\cV_{a,b})= I^{(0)}\Bigl(\cM,\cV_{a,b}\Bigr),\quad
 I_{\cN=2}(X,\cV^{bulk}_{a,b})= I^{(0)}\Bigl(X,\cV^{bulk}_{a,b}\Bigr).$$
The characters for vector bundles on $\cM$ and $X$ are computed in
Section 3.1 and Section 3.2. We use these characters in Section 3.3
to obtain the bubbled contribution and read off the unknown electric
weights $\nu_j$ and signs $s_j$ in (\ref{goal}).

 \subsection{$\mathbf{I}_{\cN=2}\bigl(\cM,\cV_{a,b}\bigr)$ and $\mathbf{I}_{\cN=4}\bigl(\cM,\cV_{a,b}\bigr)$}
In Appendix A we applied the Leray spectral sequence to compute, for any
pull-back bundle $\cV=\pi^*\bfV$ on $\cM$,
\be \label{res_des}\mathbf{I}_{\cN=4}\bigl(\cM,\cV\bigr)=\sum_{p,q}(-)^{p+q}H^p\Bigl(\cM, \Omega_{\cM}^q\otimes \cV \Bigr)=3\sum_{q=0}^2
(-)^q\hat{\chi}\Bigl(\Omega_{\PP^2}^q\otimes \bfV\Bigr)\ee
$$\mathbf{I}_{\cN=2}\bigl(\cM,\cV\bigr)=\sum_{p}(-)^{p}H^p\Bigl(\cM,\cV \Bigr)=
\hat{\chi}\Bigl(\bfV\Bigr),$$
where $\hat \chi$ denotes the weighted  sum 
$$\hat{\chi}(V):=\sum_p (-)^p H^p(\PP^2, V)$$
of bundle cohomologies on $\PP^2$.

Let us first compute $\mathbf{I}_{\cN=4}\bigl(\cM,\cV_{-1,-1}\bigr)=
\mathbf{I}_{\cN=4}\bigl(\cM,\cV_{1,1}\bigr)$ and
$\mathbf{I}_{\cN=2}\bigl(\cM,\cV_{-1,-1}\bigr)$ for 
$$\cV_{-1,-1}=\pi^*\bfV_{-1,-1}, \quad \bfV_{-1,-1}=\cO_{\PP^2}(3)\otimes \tV_1.$$
The properties (\ref{info}) allow us to identify $\tV_1$ with the vector bundle that
fits into the following exact sequence
\be \label{ex}0\mapsto \cO_{\PP^2}\mapsto \tV_1\mapsto I_p\mapsto 0,\ee
where $I_p$ is the ideal sheaf of a point on $\PP^2.$ It is useful to
recall that $I_p$ fits into another exact sequence
\be \label{ex3}0\mapsto I_p\mapsto
\cO_{\PP^2}\mapsto \cO_p\mapsto 0,\ee
where $\cO_p$ is the skyscraper sheaf supported on $p.$
We further tensor (\ref{ex}) and (\ref{ex3}) by $\cO_{\PP^2}(3)\otimes \Omega^j_{\PP^2}$ and use the corresponding long exact sequences for cohomology
groups together with Kodaira-Serre duality to find:
\be \label{zero}H^0(\PP^2,\bfV_{-1,-1})=2\mathbb{V}_{(3,0)}-\mathbb{V}_{(0,0)},\quad
H^0(\PP^2,\Omega^1_{\PP^2}\otimes\bfV_{-1,-1})=2\mathbb{V}_{(1,1)}-2\mathbb{V}_{(0,0)}\ee
\be \label{double}H^0(\PP^2,\Omega_{\PP^2}^2\otimes \bfV_{-1,-1})=\mathbb{V}_{(0,0)},\quad
H^p(\PP^2,\Omega_{\PP^2}^q\otimes \bfV_{-1,-1})=0\quad p=1,2 \,\,\, \forall q.\ee
Here $\mathbb{V}_{(m,n)}$ is an irreducible representation of $PSU(3)$ with the highest weight $(m,n).$
In total
\be \label{res0}\boxed{\mathbf{I}_{\cN=2}\Bigl(\cM,\cV_{-1,-1}\Bigr)=\hat \chi\Bigl(\bfV_{-1,-1}\Bigr)=2\VV_{(3,0)}
-\VV_{(0,0)}}\ee
\be \label{res1}\boxed{\mathbf{I}_{\cN=4}\Bigl(\cM,\cV_{1,1}\Bigr)=\mathbf{I}_{\cN=4}\bigl(\cM,\cV_{-1,-1}\bigr)=3\sum_{q=0}^2(-1)^q\hat{\chi}\Bigl(\Omega^q_{\PP^2}\otimes \bfV_{1,1}\Bigr)=6\Bigl(\mathbb{V}_{(3,0)}-\mathbb{V}_{(1,1)}+\mathbb{V}_{(0,0)}\Bigr).}\ee

Similarly, for $\bfV_{-2,-2}=\cO_{\PP^2}(6)\otimes S^2\tilde{V}_1$ we use
$$\hat \chi\Bigl(\Omega_{\PP^2}^{p}\otimes \bfV_{-2,-2}\Bigr)=\hat \chi \Bigl(\Omega_{\PP^2}^{p}\otimes \bfV_{-1,-1}\otimes \bfV_{-1,-1}\Bigr)-
\hat \chi\Bigl(\Omega_{\PP^2}^{p}\otimes \cO_{\PP^2}(6)\Bigr)\quad p=0,1,2$$

to compute
$${\hat \chi}\Bigl(\PP^2,\bfV_{-2,-2}\Bigr)=3\mathbb{V}_{(6,0)}-4\mathbb{V}_{(0,0)},
\quad {\hat \chi}\Bigl(\PP^2,\Omega^1_{\PP^2}\otimes \bfV_{-2,-2}\Bigr)=3\mathbb{V}_{(4,1)}-8\mathbb{V}_{(0,0)},$$
$$ {\hat \chi}\Bigl(\PP^2,\Omega^2_{\PP^2}\otimes \bfV_{-2,-2}\Bigr)=3\mathbb{V}_{(3,0)}-4\mathbb{V}_{(0,0)}
$$
so that
\be \label{res3}\boxed{\mathbf{I}_{\cN=2}\bigl(\cM,\cV_{-2,-2}\bigr)=3\VV_{(6,0)}
-4\VV_{(0,0)}}\ee
\be \label{res4}\boxed{\mathbf{I}_{\cN=4}\bigl(\cM,\cV_{2,2}\bigr)=\mathbf{I}_{\cN=4}\bigl(\cM,\cV_{-2,-2}\bigr)=9\Bigl(\mathbb{V}_{(6,0)}-\mathbb{V}_{(4,1)}+\mathbb{V}_{(3,0)}\Bigr).}\ee
Analogously, both the Euler and the holomorphic Euler characters can be computed straightforwardly for any $\cV_{-a,-b}=\pi^*\mathbf{V}_{-a,-b}$ with $a>0,\, b>0,\, a+2b=0 \,\, \text{mod}\, \, 3$ by
using that cohomology of the vector bundle $\mathbf{V}_{-a,-b}=\cO_{\PP^2}\bigl(2a+b\bigr)\otimes S^{b} \tilde V_1$ are determined by an iterative procedure
 \be \label{iter}
 \hat \chi \Bigl(\Omega_{\PP^2}^{p}\otimes \bfV_{-a,-b}\Bigr)=\hat \chi\Bigl(\Omega_{\PP^2}^{p}\otimes \bfV_{-(a-1),-(b-1)}\otimes \bfV_{-1,-1}\Bigr)
 - \hat \chi \Bigl(\Omega_{\PP^2}^{p}\otimes \bfV_{-(a-2),-(b-2)}\otimes \cO_{\PP^2}(6) \Bigr).\ee
For example, in this way one finds
\be \label{general}\mathbf{I}_{\cN=4}\Bigl(\cM, \cV_{a, b}\Bigr)=
\mathbf{I}_{\cN=4}\Bigl(\cM, \cV_{-a, -b}\Bigr)=3(b+1)\Bigl(\VV_{(n,0)}-\VV_{(n-2,1)}
+\VV_{(n-3,0)}\Bigr)\quad n=2a+b.\ee

\subsection{$\mathbf{I}_{\cN=2}\bigl(X,\cV^{bulk}_{a,b}\bigr)$ and 
$\mathbf{I}_{\cN=4}\bigl(X,\cV^{bulk}_{a,b}\bigr)$}

Let us denote $\pi^{bulk}:X \mapsto \PP^2.$ In this section
we first compute
$\mathbf{I}_{\cN=4}\bigl(X,\cV^{bulk}_{-1,-1}\bigr)$ with
$\cV^{bulk}_{-1,-1}=\bigl(\pi^{bulk}\bigr)^*\bfV_{-1,-1}.$

Using $\mathcal{G}_X$ defined in (\ref{def_con}), we find the norm of a section 
$\psi_{unit}=\mathcal{G}_X \begin{pmatrix} \psi_1 \cr \psi_2 \end{pmatrix}$
of the vector bundle $\cV^{bulk}_{-1,-1}$:
\be \label{norm}\vert\vert \psi_{unit}\vert\vert^2=\int {\bigl(h^{(3)}(\ts)\bigr)^2 \over y^3}\, vol_X\,\,
\Biggl(\alpha^2\vert \psi_1+\psi_2\vert^2+
\alpha^{-2}\vert \beta(\psi_1+\psi_2)+\psi_2\vert^2\Biggr)\ee
with volume form on $X$ given by
$$vol_{X}=\tf_1(\ts)\tf_2(\ts)\tf_3(\ts)\tf_4(\ts)e_1\wedge e_2\wedge e_3\wedge e_4 \wedge \overline{e}_1\wedge  \overline{e}_2\wedge
\overline{e}_3\wedge \overline{e}_4.$$
This allows to think about $\cV_{-1,-1}^{bulk}$ as follows
\be \label{def1}0\mapsto \cO_X(3) \mapsto \cV_{-1,-1}^{bulk}\mapsto I_{Z}(3)\mapsto 0\ee
\be \label{def2}0\mapsto I_Z(3)\mapsto \cO_X(3)\mapsto \cO_Z(3)\mapsto 0,\ee
where $\cO_Z$ is the structure sheaf of the fiber $Z$ of $X$ at the point $p.$

Now we note 
\be \label{simple}\mathbf{I}_{\cN=4}\Bigl(X,\cV^{bulk}_{-1,-1}\Bigr)=
2\mathbf{I}_{\cN=4}\Bigl(X,\cO_X(3)\Bigr)-\mathbf{I}_{\cN=4}\Bigl(X,\cO_Z(3)\Bigr)\ee
\be \label{simple2}\mathbf{I}_{\cN=2}\Bigl(X,\cV^{bulk}_{-1,-1}\Bigr)=
2\mathbf{I}_{\cN=2}\Bigl(X,\cO_X(3)\Bigr)-
\mathbf{I}_{\cN=2}\Bigl(X,\cO_Z(3)\Bigr)\ee
To evaluate $\mathbf{I}_{\cN=4}\Bigl(X,\cO_Z(3)\Bigr)$ and $\mathbf{I}_{\cN=2}\Bigl(X,\cO_Z(3)\Bigr)$, we use the fact that the only\footnote{Higher cohomology groups vanish since $Z$ is topologically $\CC^2$} non-zero cohomology groups
involving $\cO_Z(3),$  are
\be \label{nonzero}
H^0(X,\cO_Z(3))=3\VV_{(0,0)},\quad H^0(X,\Omega^1_X \otimes \cO_Z(3))=8\VV_{(0,0)},
\quad H^0(X,\Omega^2_X \otimes \cO_Z(3))=\VV_{(0,0)}\ee
Indeed, we found the following non-zero cohomology groups:
\begin{itemize}
\item $H^0\Bigl(X,\cO_X(3)\Bigr)=\mathbb{V}_{(3,0)}
+\mathbb{V}_{(1,1)}$
\item $H^0\Bigl(X,\Omega^1_X(3)\Bigr)=2\mathbb{V}_{(1,1)}+\VV_{(0,0)}$
\item $H^0\Bigl(X,\Omega^2_{X}(3)\Bigr)=
\mathbb{V}_{(0,0)}$
\end{itemize}
Explicitly,  sections in  $H^0\Bigl(X,\cO_X(3)\Bigr)$ (in the holomorphic gauge)
are
\begin{enumerate}
\item $M^i_jt_iU^j \begin{pmatrix}1 \cr -1 \end{pmatrix} \,\,\, i,j=1,\ldots 3\,\,\, \text{with}\,\, M^i_i=0$
transform in $\mathbb{V}_{(1,1)}$ 
\item $C_{ijk}U^iU^jU^k $ transform in $\VV_{(3,0)}$
\end{enumerate}

Meanwhile, sections in $H^0\Bigl(X,\Omega^1_X(3)\Bigr)$
are (in the holomorphic gauge)

\begin{enumerate}
\item $h^i_k(U^kdt_j-2t_jdU^k)$
transform in $\mathbb{V}_{(1,1)}+\VV_{(0,0)}.$
\item $C_{i[jk]}U^iU^jdU^k$  transform in $\VV_{(1,1)}.$
\end{enumerate}

Finally, $H^0\Bigl(X,\Omega^2_X(3)\Bigr)$ is one-dimensional and generated by (in the holomorphic gauge)
$$\epsilon_{ijk}U^idU^j\wedge dU^k.$$
In (\ref{nonzero}) we simply count cohomology elements which are non-vanishing at $z^1=z^2=0.$
So we compute
$$\mathbf{I}_{\cN=4}\Bigl(X,\cO_Z(3)\Bigr)=-4\VV_{(0,0)},\quad
\mathbf{I}_{\cN=2}\Bigl(X,\cO_Z(3)\Bigr)=3\VV_{(0,0)}.$$
Meanwhile, the characters for $\cO_X(3)$ are computed in Appendix B:
$$\boxed{\mathbf{I}_{\cN=2}\Bigl(X,\cO_X(3)\Bigr)=\VV_{(3,0)}+\VV_{(1,1)}}\quad
\boxed{\mathbf{I}_{\cN=4}\Bigl(X,\cO_X(3)\Bigr)=\VV_{(3,0)}-\VV_{(1,1)}+\VV_{(0,0)}}$$
Hence, we find the Euler and the holomorphic Euler characters of $\cV^{bulk}_{-1,-1}$
$$\boxed{\mathbf{I}_{\cN=4}\Bigl(X,\cV^{bulk}_{1,1}\Bigr)=\mathbf{I}_{\cN=4}\Bigl(X,\cV^{bulk}_{-1,-1}\Bigr)=
2\Biggl(\VV_{(3,0)}-\VV_{(1,1)}+\VV_{(0,0)}\Biggr)+4\VV_{(0,0)}}$$
$$\boxed{\mathbf{I}_{\cN=2}\Bigl(X,\cV^{bulk}_{-1,-1}\Bigr)=2\Bigl(\VV_{(3,0)}+\VV_{(1,1)} \Bigr)-3\VV_{(0,0)}}.$$

Let us consider $\cV^{bulk}_{-2,-2}$ as another example. We compute 
$$I^{\a}\Bigl(X,\cV^{bulk}_{-2,-2}\Bigr)=I^{\a}\Bigl(X,\cV^{bulk \, \otimes 2}_{-1,-1}\Bigr)-
I^{\a}\Bigl(X,\cO_X(6)\Bigr)$$
Next we use, in addition to (\ref{def1}) and (\ref{def2}), the following short exact sequence
\be \label{def3}
0 \mapsto \cO_X(3)\otimes \cV_{-1,-1}^{bulk} \mapsto \cV_{-1,-1}^{bulk \, \otimes 2}\mapsto 
I_{Z}(3)\otimes \cV_{-1,-1}^{bulk}\mapsto 0\ee

So that
\be \label{answ} \boxed{I^{\a}\Bigl(X,\cV^{bulk}_{-2,-2}\Bigr)=
3I^{\a}\Bigl(X,\cO_X(6)\Bigr)
-3I^{\a}\Bigl(X,\cO_Z(6)\Bigr)}\ee
where we used
\be \label{vip}I^{\a}\Bigl(X,\cV^{bulk}_{-1,-1}(3)\otimes \cO_Z\Bigr)=I^{\a}\Bigl(X,\cO_Z(6)\Bigr).\ee
We compute $I^{\a}\Bigl(X,\cO_Z(6)\Bigr)$ by first listing
all generators of $\HH^0\Bigl(X,\Omega^{\alpha}_X(6)\Bigr)$ and then
picking up those which do not vanish at $z_1=z_2=0.$ 
Only 23 of these are non-vanishing at $Z$ so that 
$$I^{(1)}\Bigl(X,\cO_Z(6)\Bigr)=23\VV_{(0,0)}.$$

Similarly we computed
$$I^{(0)}\Bigl(X,\cO_Z(6)\Bigr)=10\VV_{(0,0)},\,\,
I^{(2)}\Bigl(X,\cO_Z(6)\Bigr)=14\VV_{(0,0)},\,\,
I^{(3)}\Bigl(X,\cO_Z(6)\Bigr)=11\VV_{(0,0)},\,\, I^{(4)}\Bigl(X,\cO_Z(6)\Bigr)=0$$
so that
$$\boxed{\mathbf{I}_{\cN=4}\Bigl(X,\cO_Z(6)\Bigr)=-10\VV_{(0,0)},\quad
\mathbf{I}_{\cN=2}\Bigl(X,\cO_Z(6)\Bigr)=10\VV_{(0,0)}}$$
Using $\mathbf{I}_{\cN=4}\Bigl(X,\cO_X(6)\Bigr)$ computed in Appendix B, we
finally get
\be \label{res}\boxed{\mathbf{I}_{\cN=4}\Bigl(X,\cV_{2,2}^{bulk}\Bigr)=\mathbf{I}_{\cN=4}\Bigl(X,\cV_{-2,-2}^{bulk}\Bigr)=3\Biggl(\VV_{(6,0)}-\VV_{(4,1)}+
\VV_{(0,3)}+\VV_{(3,0)}-\VV_{(1,1)}\Biggr)+30\VV_{(0,0)}}\ee
\be \label{resii}\boxed{\mathbf{I}_{\cN=2}\Bigl(X,\cV_{-2,-2}^{bulk}\Bigr)=
3\Biggl(\VV_{(6,0)}+\VV_{(4,1)}+\VV_{(2,2)}+
\VV_{(0,3)}\Biggr)-30\VV_{(0,0)}}\ee
One may use an iterative procedure to write down the characters for general $\cV^{bulk}_{-a,-b}$:
\be\label{iter_ii}I^{\a}\Bigl(X, \cV^{bulk}_{-a,-b}\Bigr)=2I^{\a}\Bigl(X, \cV^{bulk}_{-(a-1), -(b-1)}\otimes \cO_X(3)\Bigr)-I^{\a}\Bigl(X, \cV^{bulk}_{-(a-1), -(b-1)}\otimes \cO_Z(3)\Bigr)\ee
$$-I^{\a}\Bigl(X, \cV^{bulk}_{-(a-2), -(b-2)}\otimes \cO_X(6)\Bigr)\quad \a=0,\ldots,4.$$
\subsection{Finding the OPE}
To determine the unknown electric weights $\nu_j$ and signs $s_j$
on the right side of the OPE (\ref{goal}), we need to decompose the bubbled contributions 
$${\bf I}^{bubble}_{\cN=2}\Bigl(\cV_{a,b}\Bigr)={\bf I}_{\cN=2}\Bigl(\cM,\cV_{a,b}\Bigr)-
{\bf I}_{\cN=2}\Bigl(X,\cV^{bulk}_{a,b}\Bigr)$$
and
$${\bf I}^{bubble}_{\cN=4}\Bigl(\cV_{a,b}\Bigr)={\bf I}_{\cN=4}\Bigl(\cM,\cV_{a,b}\Bigr)-
{\bf I}_{\cN=4}\Bigl(X,\cV^{bulk}_{a,b}\Bigr)$$
into representations of the $SU(2)_{\alpha_1}\times U(1)$ which is unbroken in the presence
of 't Hooft operator with magnetic charge $w_2.$
For example, for
\be \label{imp_new}{\bf I}^{bubble}_{\cN=2}\Bigl(\cV_{-1,-1}\Bigr)=2\Bigl( \VV_{(0,0)}-\VV_{(1,1)}\Bigr)\quad {\bf I}^{bubble}_{\cN=4}\Bigl(\cV_{-1,-1}\Bigr)={\bf I}^{bubble}_{\cN=4}\Bigl(\cV_{1,1}\Bigr)=4\Bigl(\VV_{(3,0)}-\VV_{(1,1)}+\VV_{(0,0)}\Bigr)+2\VV_{(0,0)}\ee
we use
$$\VV_{(3,0)}\mapsto \mathbf{R}^{(d=4)}_{(3,0)}+\mathbf{R}^{(d=2)}_{(1,-2)}+
\mathbf{R}^{(d=3)}_{(2,-1)}+\mathbf{R}^{(d=1)}_{(0,-3)}$$
$$\VV_{(1,1)}\mapsto \mathbf{R}^{(d=2)}_{(1,1)}+\mathbf{R}^{(d=2)}_{(1,-2)}+
\mathbf{R}^{(d=3)}_{(2,-1)}+\mathbf{R}^{(d=1)}_{(0,0)}$$
where $\mathbf{R}^{(d)}_{(m,n)}$ stands for  $d$-dimensional representation
with the highest weight $(m,n)$ and all other vectors in this representations are obtained
by acting with the lowering operator $-\alpha_1.$
So that we can bring (\ref{imp_new}) into the form suitable for reading off the answer
for the OPE:
\be \label{prelim}{\bf I}^{bubble}_{\cN=2}\Bigl(\cV_{-1,-1}\Bigr)=-2\Bigl(\mathbf{R}^{(d=2)}_{(1,1)}+\mathbf{R}^{(d=2)}_{(1,-2)}+
\mathbf{R}^{(d=3)}_{(2,-1)}\Bigr)\ee
\be \label{prelim_ii}{\bf I}^{bubble}_{\cN=4}\Bigl(\cV_{-1,-1}\Bigr)={\bf I}^{bubble}_{\cN=4}\Bigl(\cV_{1,1}\Bigr)=4\Bigl(\mathbf{R}^{(d=4)}_{(3,0)}-\mathbf{R}^{(d=2)}_{(1,1)}+\mathbf{R}^{(d=1)}_{(0,-3)}\Bigr) +2\mathbf{R}^{(d=1)}_{(0,0)}.\ee

From (\ref{prelim}) we obtain  the following OPE in $\cN=2$ SYM with gauge group $G=PSU(3)$:
\be \label{main_i}\boxed{WT_{w_1;-w_1-w_2}\times WT_{w_1;0}=
WT_{2w_1;-w_1-w_2}-2\Bigl(WT_{w_2;w_1+w_2}+WT_{w_2;w_1-2w_2}+
WT_{w_2;2w_1-w_2}\Bigr)}\ee
Meanwhile, in $\cN=4$ SYM with gauge group $G=PSU(3)$
we find from (\ref{prelim_ii}):
\be \label{main_ii}\boxed{WT_{w_1;-w_1-w_2}\times WT_{w_1;0}=
WT_{2w_1;-w_1-w_2}+4\Bigl(WT_{w_2;3w_1}-WT_{w_2;w_1+w_2}+
WT_{w_2;-3w_2)}\Bigr)+2WT_{w_2; 0}}\ee
\be \label{main_iii}\boxed{WT_{w_1;w_1+w_2}\times WT_{w_1;0}=
WT_{2w_1;w_1+w_2}+4\Bigl(WT_{w_2;3w_1}-WT_{w_2;w_1+w_2}+
WT_{w_2;-3w_2)}\Bigr)+2WT_{w_2; 0}}\ee
Similarly, for
\be \label{imp2}{\bf I}^{bubble}_{\cN=2}\Bigl(\cV_{-2,-2}\Bigr)=26\VV_{(0,0)}-3\Bigl( \VV_{(4,1)}+\VV_{(2,2)}+\VV_{(0,3)}\Bigr)\ee
\be \label{imp3} {\bf I}^{bubble}_{\cN=4}\Bigl(\cV_{-2,-2}\Bigr)={\bf I}^{bubble}_{\cN=4}\Bigl(\cV_{2,2}\Bigr)=6\Bigl(\VV_{(6,0)}-\VV_{(4,1)}+\VV_{(3,0)}\Bigr)-30\VV_{(0,0)}-3\Bigl(\VV_{(0,3)}-\VV_{(1,1)}\Bigr)\ee
we use

$$\VV_{(2,2)}\mapsto \mathbf{R}^{(d=5)}_{(4,-2)}+\mathbf{R}^{(d=4)}_{(3,-3)}+
\mathbf{R}^{(d=4)}_{(3,0)}+\mathbf{R}^{(d=3)}_{(2,2)}+
\mathbf{R}^{(d=3)}_{(2,-4)}+
\mathbf{R}^{(d=3)}_{(2,-1)}+\mathbf{R}^{(d=2)}_{(1,-2)}
+\mathbf{R}^{(d=2)}_{(1,1)}+\mathbf{R}^{(d=1)}_{(0,0)}$$
$$\VV_{(0,3)}\mapsto \mathbf{R}^{(d=4)}_{(3,-3)}+
\mathbf{R}^{(d=3)}_{(2,-1)}+\mathbf{R}^{(d=2)}_{(1,1)}+\mathbf{R}^{(d=1)}_{(0,3)}$$
$$\VV_{(4,1)}\mapsto \mathbf{R}^{(d=6)}_{(5,-1)}+\mathbf{R}^{(d=5)}_{(4,1)}+
\mathbf{R}^{(d=5)}_{(4,-2)}+\mathbf{R}^{(d=4)}_{(3,-3)}+\mathbf{R}^{(d=4)}_{(3,0)}+
\mathbf{R}^{(d=3)}_{(2,-1)}+\mathbf{R}^{(d=3)}_{(2,-4)}+\mathbf{R}^{(d=2)}_{(1,-2)}
+\mathbf{R}^{(d=2)}_{(1,-5)}+\mathbf{R}^{(d=1)}_{(0,-3)}$$
$$\VV_{(6,0)}\mapsto \mathbf{R}^{(d=7)}_{(6,0)}+\mathbf{R}^{(d=6)}_{(5,-1)}+
\mathbf{R}^{(d=5)}_{(4,-2)}+\mathbf{R}^{(d=4)}_{(3,-3)}+\mathbf{R}^{(d=3)}_{(2,-4)}+
\mathbf{R}^{(d=2)}_{(1,-5)}+\mathbf{R}^{(d=1)}_{(0,-6)}$$
to get the following OPE in $\cN=2$ SYM with gauge group $G=PSU(3)$:

\be \label{main_iv}WT_{w_1;-2w_1-2w_2}\times WT_{w_1;0}=
WT_{2w_1;-2w_1-2w_2}-3\, WT_{w_2;5w_1-w_2}-3\,WT_{w_2;4w_1+w_2}-6\,
WT_{w_2;4w_1-2w_2}\ee
$$-6\,WT_{w_2;3w_1-3w_2}-6\,WT_{w_2;3w_1}-3\,WT_{w_2;2w_1+2w_2}
-6\,WT_{w_2;2w_1-4w_2}-9\,WT_{w_2;2w_1-w_2}$$
$$-6\,WT_{w_2;w_1-2w_2}-6\,WT_{w_2;w_1+w_2}-3WT_{w_2;w_1-5w_2}-3\,
WT_{w_2;3w_2}-3\,WT_{w_2;-3w_2}+23WT_{w_2;0}$$

Meanwhile, for $\cN=4$ SYM with gauge group $G=PSU(3)$ we obtain:
\be \label{main_v}WT_{w_1; -2w_1-2w_2}\times WT_{w_1; 0}=
WT_{2w_1; -2w_1-2w_2}+6\, WT_{w_2; 6w_1}-6\, WT_{w_2; 4w_1+w_2}+
6\, WT_{w_2; -6w_2}\ee
$$-3\, WT_{2w_1; 3w_1-3w_2}+3\,  WT_{2w_1; w_1-2w_2}
-3\, WT_{2w_1; 3w_2}-27\, WT_{2w_1; 0}$$

\be \label{main_vi}WT_{w_1; 2w_1+2w_2}\times WT_{w_1; 0}=
WT_{2w_1; 2w_1+2w_2}+6\, WT_{w_2; 6w_1}-6\, WT_{w_2; 4w_1+w_2}+
6\, WT_{w_2; -6w_2}\ee
$$-3\, WT_{2w_1; 3w_1-3w_2}+3\,  WT_{2w_1; w_1-2w_2}
-3\, WT_{2w_1; 3w_2}-27\, WT_{2w_1; 0}$$

\section{Conclusion}
In this paper we computed the OPE (\ref{goal}) of Wilson-'t Hooft loop operators
in $\cN=4$ and $\cN=2$ SYM theory with gauge group $G=PSU(3).$
This work is an extension of our approach \cite{KS},\cite{Saulina} which
uses the holomorphic-topological twist  \cite{htft} of
the $\cN=2$ gauge theory and the connection between BPS configurations  
in the presence of 't Hooft operators and solutions of 3d Bogomolny equations 
with magnetic sources \cite{KW},\cite{Witten2009}.

The crucial ingredients in our computation of the OPE in $\cN=4$ SYM are
Euler characters ${\bf I}_{\cN=4}\Bigl(\cM,\cV_{a,b}\Bigr)$ and ${\bf I}_{\cN=4}\Bigl(X,\cV^{bulk}_{a,b}\Bigr)$ of vector bundles $\cV_{a,b}$ and $\cV^{bulk}_{a,b}$
 on moduli spaces
$\cM$ and $X=T\PP^2.$ These Euler characters compute (with sign) the
ground states of the supersymmetric quantum mechanics which arises as 
the result of quantizing  $\cN=4$ SYM on $\RR\times I\times \mathcal{C}$ in the presence of Wilson-t' Hooft operators along $\RR.$ To find the OPE
in $\cN=2$ SYM, we compute the holomorphic Euler characters instead of
Euler characters relevant for $\cN=4$ SYM.

In Section 3, we explained how to compute
the right side of (\ref{goal}) for any $\nu=aw_1+bw_2.$ Namely, we
used that $\cM$ is $\PP^2$ fibration over $\PP^2$ and applied the Leray
spectral sequence in Appendix A.  Together with the iterative procedure (\ref{iter}) for vector bundles on $\PP^2$, this allows to reduce the problem of finding 
${\bf I}_{\cN=4}\Bigl(\cM,\cV_{a,b}\Bigr)$ to 
computing the Euler characters
of line bundles on $\PP^2.$ Further, due to the iterative procedure (\ref{iter_ii})
for vector bundles on $X=T\PP^2$, 
the problem of
finding ${\bf I}_{\cN=4}\Bigl(X,\cV^{bulk}_{a,b}\Bigr)$ is reduced to
computing the Euler character of $L^2$ Dolbeault cohomology of line bundles on $X.$ This step is done in Appendix B by using the exact sequences for $\Omega^{\a}_X$ for $\a=1,\ldots,4.$ Finally, to get the explicit answers for the OPE,
given in equations (\ref{main_i}-\ref{main_iii}) and
(\ref{main_iv}-\ref{main_vi}), we used vanishing assumption (\ref{vanish}).
We provided evidence in support of this assumption in Appendix C. 

\section*{Acknowledgments}

We would like to thank C. Cordova, S. Gukov, G. Moore, A. Neitzke, A. Kapustin, S. Katz, ~~~C. Vafa, E. Witten  for
discussions. N. S. is grateful to Simons Center for Geometry and Physics and
Institute for Advanced Study for the hospitality at various stages of this work.

 \startappendix
 \section{Computing $\mathbf{I}_{\cN=4}\Bigl(\cM,\cV\Bigr)$}
 Here we determine
 $$\mathbf{I}_{\cN=4}\bigl(\cM,\cV\bigr)=\sum_{p,q}(-)^{p+q}H^p\Bigl(\cM, \Omega_{\cM}^q\otimes \cV \Bigr).$$
 Let $\pi: \cM \mapsto \PP^2$ and $\cV=\pi^* \mathbf{V}$ where $\mathbf{V}$
 is a vector bundle on $\PP^2.$
 To compute cohomology $H^j(\cM, \Omega_{\cM}^n\otimes \mathcal{V})$ we use the Leray spectral
 sequence.  This is easy enough since we are interested in pull-back bundles.
 The necessary ingredients are the right direct images of $\Omega_{\cM}$ and
 $\Omega^2_{\cM}$ which we compute in sections A.1 and A.2 respectively.
 
\subsection{ $R^p\pi_*\Omega_{\cM}$}
The 4-fold $\cM$ is defined by $y_aU^a=0$ in the toric 5-fold $Y_5.$ 
The weights under the two $\mathbb{C}^*$ actions are
\be \label{action}\begin{array}{c|c|c|c|c|c|c|c}
&U^1 & U^2 & U^3 & \Lambda & y_1 & y_2 & y_3\cr
\hline
\text{new}&1 & 1 & 1& -2& 0& 0& 0 \cr
\hline
\text{old}&0& 0& 0& 1& 1&1&1\cr
\end{array}
\ee
Here $U^1,U^2,U^3$ are homogenous coordinates on the base $\mathbb{P}_{\vec U}^2$
and $\Lambda, y_1,y_2,y_3$  are homogenous coordinates on the
fiber $\mathbb{P}^3.$ 

Let us use that $Y_5$ is the projectivisation of the vector bundle
$$Y_5=\PP(E),\quad E=\cO_{\PP^2}(-2)\oplus\cO_{\PP^2}^{\oplus 3}$$
to write the following two exact sequences:
\be \label{exi}0\mapsto \hat{\pi}^*\Omega^1_{\PP^2}\mapsto \Omega^1_{Y_5}\mapsto
\Omega^1_{vert}\mapsto 0\ee
\be \label{exii}0\mapsto \Omega^1_{vert}\mapsto \hat{\pi}^*E^*\otimes \cO_{Y_5}(0,-1)\mapsto \cO_{Y_5}\mapsto 0.\ee
Here and below $\cO_{Y_5}(b,f)$ stands for a line bundle on $Y_5$
with degree $b$ on the base $\PP^2$ and degree $f$ on the fiber $\PP^3.$
More explicitly, the exact sequence (\ref{exii}) is
\be\label{exiii}0\mapsto \Omega^1_{vert}\mapsto \cO_{Y_5}(2,-1)\oplus \cO_{Y_5}(0,-1)^{\oplus 3}\mapsto \cO_{Y_5}\mapsto 0.\ee
Let us denote
$$\pi: \cM \mapsto \mathbb{P}^2,\quad \hat{\pi}: Y_5 \mapsto \mathbb{P}^2.$$
We apply push-forward map to (\ref{exiii}) and use 
$$H^i\Bigl(\PP^3, \cO_{\PP^3}(-1)\Bigr)=0\quad \text{for}\quad  i=0,\ldots,3, \quad
H^0\Bigl(\PP^3, \cO_{\PP^3}\Bigr)=\CC,\quad H^i\Bigl(\PP^3, \cO_{\PP^3}\Bigr)=0\quad \text{for}\quad i>0$$
to compute the direct images of $\Omega^1_{vert}$:
\be \label{dir}R^1\hat{\pi}_*\Bigl(\Omega^1_{vert}\Bigr)=\cO_{\PP^2},\quad
R^k\hat{\pi}_*\Bigl(\Omega^1_{vert}\Bigr)=0 \quad \text{for} \quad k\ne 1.\ee
Now we apply the push-forward map to (\ref{exi}) to
find
\be \label{dir_res}\hat{\pi}_*\Bigl(\Omega^1_{Y_5}\Bigr)=\Omega^1_{\PP^2},\quad
R^1\hat{\pi}_*\Bigl(\Omega^1_{Y_5}\Bigr)=\cO_{\PP^2},\quad
R^k\hat{\pi}_*\Bigl(\Omega^1_{Y_5}\Bigr)=0 \quad \text{for} \quad k>1.
\ee

Next we use the adjunction formula
$$0\mapsto N^*_{\cM\vert Y_5}\mapsto \Omega^1_{Y_5}\vert_{\cM}\mapsto
\Omega^1_{\cM}\mapsto 0$$
where $N^*_{\cM\vert Y_5}=\mathcal{O}_{\cM}(-1,-1)$ is the co-normal bundle
of $\cM$ in $Y_5.$

Now we use
$R^i \pi_*i^*=R^i\hat{\pi}_*$ and $H^p\bigl(\mathbb{P}^2,
\mathcal{O}_{\mathbb{P}^2}(-1)\bigr)=0$ for all $p$ to
find
$$R^i\pi_*\Omega^1_{\cM}=R^i\hat{\pi}_*\Omega^1_{Y_5}.$$
Therefore, we conclude
\be \label{imp}\pi_*\Omega^1_{\cM}=\Omega^1_{\mathbb{P}^2},\quad
R^1\pi_*\Omega^1_{\cM}=\mathcal{O}_{\mathbb{P}^2},\quad
R^p \pi_*\Omega^1_{\cM}=0 \,\, \, p>1.\ee
We use these push-forwards to compute $E_2^{pq}=E_{\infty}^{pq}$ with $p+q=n$
giving
$$H^1(\cM,\Omega^1_{\cM})=\mathbb{V}_1+\mathbb{V}_1,\quad \quad
H^p(\cM,\Omega^1_{\cM})=0 \,\,\, p\ne 1.$$
This agrees with the cohomology of $\cM$ computed before in a different way.

\subsection{$R^p\pi_*\Omega^2_{\cM}$}
We first use three short exact sequences:
\begin{enumerate}
\item $0\mapsto \mathcal{N}^*_{\cM\vert Y_5}\otimes \Omega^1_{\cM}\mapsto \Omega^2_{Y_5}\vert_{\cM}\mapsto \Omega^2_{\cM}\mapsto 0$

\item $0\mapsto \Omega^1_{\mathbb{P}^2}\mapsto \mathcal{O}(-1)^{\oplus 3}\mapsto
\mathcal{O}\mapsto 0$

\item $0\mapsto \mathcal{O}^{\oplus 2}\mapsto
\Omega^1_{\cM\vert_{\mathbb{P}_{f}^2}}\mapsto
\Omega^1_{\mathbb{P}^2_{f}}\mapsto 0$
\end{enumerate}
to compute
$$R^p\pi_*\bigl(\mathcal{N}^*_{\cM\vert Y_5}\otimes \Omega^1_{\cM}\bigr)=0\quad p=0,1,2$$
and
\be \label{answer} R^p\hat{\pi}\bigl(\Omega_{Y_5}^2\bigr)=R^p\pi_*\bigl(\Omega_{\cM}^2\bigr)\quad p=0,1,2.\ee

Now we use the following filtration on $Y_5$:
$$0 \subset \mathbb{F}_1\subset \mathbb{F}_2\subset \mathbb{F}_3=\Omega^2_{Y_5}$$
Here
$$\mathbb{F}_1=\Lambda^2\Bigl(\hat{\pi}^*\Omega^1_{\PP^2}\Bigr),\quad
\mathbb{F}_2/\mathbb{F}_1=\Bigl(\hat{\pi}^*\Omega^1_{\PP^2}\Bigr)\otimes
\Omega^1_{vert},\quad \mathbb{F}_3/\mathbb{F}_2=\Lambda^2\Omega^1_{vert}.$$
From the push-forward of the following exact sequence
$$0\mapsto \mathbb{F}_1 \mapsto \mathbb{F}_2 \mapsto \mathbb{F}_2/\mathbb{F}_1\mapsto 0$$
we find
\be \label{res_ii}\hat{\pi}_*\Bigl(\mathbb{F}_2\Bigr)=\Omega^2_{\PP^2},\quad
R^1\hat{\pi}_*\Bigl(\mathbb{F}_2\Bigr)=\Omega^1_{\PP^2},\quad
R^j\hat{\pi}_*\Bigl(\mathbb{F}_2\Bigr)=0 \,\,\, j>1.\ee
Then, from the push-forward of the other exact sequence
$$0\mapsto \mathbb{F}_2 \mapsto \mathbb{F}_3 \mapsto \mathbb{F}_3/\mathbb{F}_2\mapsto 0$$
we compute
\be \label{res_iii}\hat{\pi}_*\Bigl(\mathbb{F}_3\Bigr)=\Omega^2_{\PP^2},\quad
R^1\hat{\pi}_*\Bigl(\mathbb{F}_3\Bigr)=\Omega^1_{\PP^2},\quad
R^2\hat{\pi}_*\Bigl(\mathbb{F}_3\Bigr)=\cO_{\PP^2},\quad
R^j\hat{\pi}_*\Bigl(\mathbb{F}_3\Bigr)=0 \,\,\, j>2.\ee

We conclude using (\ref{answer}) and (\ref{res_iii})
$$\pi_*\Bigl(\Omega^2_{\cM}\Bigr)=\Omega^2_{\PP^2},\quad
R^1\pi_*\Bigl(\Omega^2_{\cM}\Bigr)=\Omega^1_{\PP^2},\quad
R^2\pi_*\Bigl(\Omega^2_{\cM}\Bigr)=\cO_{\PP^2},\quad
R^j\pi_*\Bigl(\Omega^2_{\cM}\Bigr)=0 \,\,\, j>2.$$

As a check, we find
$$E_2^{02}=H^0(\mathbb{P}^2, R^2\pi_*\bigl(\Omega_{\cM}^2\bigr)\Bigr)=
H^0(\mathbb{P}^2, \mathcal{O})=\mathbb{V}_1$$
$$E_2^{20}=H^2(\mathbb{P}^2, \pi_*\bigl(\Omega_{\cM}^2\bigr)\Bigr)=
H^0(\mathbb{P}^2, \mathcal{O})=\mathbb{V}_1$$
$$E_2^{11}=H^1(\mathbb{P}^2, R^1\pi_*\bigl(\Omega_{\cM}^2\bigr)\Bigr)=
H^0(\mathbb{P}^2, \mathcal{O})=\mathbb{V}_1$$
giving
$$H^2\bigl(\cM, \Omega^2_{\cM}\bigr)=\mathbb{V}_1\oplus \mathbb{V}_1\oplus \mathbb{V}_1$$
in complete agreement with the analysis in \cite{Saulina}.

\subsection{$H^j\Bigl(\cM,\Omega_{\cM}^p\otimes \cV\Bigr)$}

Now we use $\cV=\pi^*(\bfV)$ and the right direct images computed in A.1 and A.2
$$\pi_*\Omega^1_{\cM}=\Omega^1_{\PP^2},\quad R^1\pi_*\Bigl(\Omega^1_{\cM}\Bigr)=\cO_{\PP^2},\quad R^j\pi_*\Bigl(\Omega^1_{\cM}\Bigr)=0 \,\, j\ge 2 $$
$$\pi_*\Bigl(\Omega^2_{\cM}\Bigr)=\Omega^2_{\PP^2},\quad
R^1\pi_*\Bigl(\Omega^2_{\cM}\Bigr)=\Omega^1_{\PP^2},\quad
R^2\pi_*\Bigl(\Omega^2_{\cM}\Bigr)=\cO_{\PP^2},\quad
R^j\pi_*\Bigl(\Omega^2_{\cM}\Bigr)=0 \,\, j\ge 3 $$
to compute bundle cohomology:

$$H^0\Bigl(\cM, \Omega_{\cM}^1\otimes \mathcal{V}\Bigr)=H^0\Bigl(\PP^2,\Omega^1_{\PP^2}\otimes \bfV \Bigr),\quad H^1\Bigl(\cM, \Omega^1_{\cM}\otimes \mathcal{V}\Bigr)=H^1\Bigl(\PP^2,\Omega^1_{\PP^2}\otimes \bfV \Bigr)+H^0\Bigl(\PP^2,\bfV \Bigr)$$
\be \label{omega}
H^2\Bigl(\cM, \Omega_{\cM}^1\otimes \mathcal{V}\Bigr)=H^2\Bigl(\PP^2,\Omega^1_{\PP^2}\otimes \bfV \Bigr)+H^1\Bigl(\PP^2,\bfV \Bigr)\ee
$$ H^3\Bigl(\cM, \Omega_{\cM}^1\otimes \mathcal{V}\Bigr)=H^2\Bigl(\PP^2,\bfV \Bigr),\quad  H^4\Bigl(\cM, \Omega_{\cM}^1\otimes \mathcal{V}\Bigr)=0,$$
and
$$H^0\Bigl(\cM, \Omega_{\cM}^2\otimes \mathcal{V}\Bigr)=H^0\Bigl(\PP^2,\Omega^2_{\PP^2}\otimes \bfV \Bigr),\quad H^1\Bigl(\cM, \Omega^2_{\cM}\otimes \mathcal{V}\Bigr)=H^1\Bigl(\PP^2,\Omega^2_{\PP^2}\otimes \bfV \Bigr)+H^0\Bigl(\PP^2,\Omega^1_{\PP^2}\otimes \bfV \Bigr)$$
\be \label{omega_ii}
H^2\Bigl(\cM, \Omega_{\cM}^2\otimes \mathcal{V}\Bigr)=
H^2\Bigl(\PP^2,\Omega^2_{\PP^2}\otimes \bfV \Bigr)+
H^1\Bigl(\PP^2,\Omega^1_{\PP^2}\otimes\bfV \Bigr)+H^0\Bigl(\PP^2,\bfV \Bigr)\ee
$$ H^3\Bigl(\cM, \Omega_{\cM}^2\otimes \mathcal{V}\Bigr)=H^1\Bigl(\PP^2,\bfV \Bigr)+H^2\Bigl(\PP^2,\Omega^1_{\PP^2}\otimes\bfV \Bigr),\quad
H^4\Bigl(\cM, \Omega_{\cM}^2\otimes \mathcal{V}\Bigr)=
H^2\Bigl(\PP^2,\bfV \Bigr).$$

Now we use Kodaira-Serre duality
$$H^n(\cM,\Omega^3_{\cM}\otimes \cV)=H^{4-n}(\cM,\Omega^1_{\cM}\otimes
\cV^*),\quad H^n(\cM,\Omega^4_{\cM}\otimes \cV)=H^{4-n}(\cM,
\cV^*)$$
For any vector bundle $V$ on $\PP^2$ let us denote the weighted  sum of
cohomologies as
$$\hat{\chi}(V):=\sum_p (-)^p H^p(\PP^2, V).$$

We now compute the desired expression
\be \label{res}\mathbf{I}_{\cN=4}\bigl(\cM,\cV\bigr)=\sum_{p,q}(-)^{p+q}H^p\Bigl(\cM, \Omega_{\cM}^q\otimes \cV \Bigr)=3\sum_{q=0}^2
(-)^q\hat{\chi}\Bigl(\Omega_{\PP^2}^q\otimes \bfV\Bigr)\ee
Note that
$$\mathbf{I}_{\cN=4}\bigl(\cM,\cV\bigr)=\mathbf{I}_{\cN=4}\bigl(\cM,\cV^*\bigr).$$

  \section{$I_{\cN=4}\Bigl(T\PP^2,\cO_{T\PP^2}(3m)\Bigr) \,\, \text{for}\,\, m\ge 1$}
  \label{appendix_one}
   \subsection{$L^2$ Dolbeault cohomology  of $\cO_{X}(m)$ with $m\ge 1$}
Recall that $X=T\PP^2$ is defined by a hypersurface $t_iU^i=0$ in $Y:=W\PP_{111222}/\{U^i=0\}.$ 
  By explicit counting of global holomorphic sections, we find
   $$\HH^0\Bigl(X,O_X(1)\Bigr)=\VV_{(1,0)},\quad
  \HH^0\Bigl(X,O_X(2)\Bigr)=\VV_{(2,0)}+\VV_{(0,1)},\quad
  \HH^0\Bigl(X,O_X(3)\Bigr)=\VV_{(3,0)}+\VV_{(1,1)}$$
  $$\HH^0\Bigl(X,O_X(4)\Bigr)=\VV_{(4,0)}+\VV_{(2,1)}+\VV_{(0,2)}$$
  More generally,
  \be \label{A1_1}\boxed{\HH^0\Bigl(X,O_X(n)\Bigr)=\sum_{j=0}^{\left[ {n\over 2} \right]} \VV_{(n-2j,j)} \quad n\ge 1}\ee
  We checked that these sections have finite norm by using the K\"ahler form on $X$
  (\ref{k_form}) and connection (\ref{con_i}-\ref{con_ii}) on line bundle $\cO_X(n).$
   
In this paper we {\it assume} the following vanishing of $L^2$ Dolbeault cohomology:
  \be \label{vanish} \boxed{\HH^j(X,O_X(n)) =0 \quad n\ge -3 \quad j>0}\ee
  For $n=0$ this was proved in \cite{Saulina},
  we checked this statement for several values of $n\ge -3$ and $j>0$ and sample computations are presented in Appendix C. 

\subsection{$L^2$ Dolbeault cohomology  of $\Omega^1_{X}(3m)$ with $m\ge 1$}
 \label{appendix_two}
 Here we express $I^{(1)}\Bigl(X,O_X(3m)\Bigr)$ for $m\ge 1$ in terms of
 holomorphic Euler characters of line bundles $I^{(0)}\Bigl(X,O_X(3m-n)\Bigr)$ for $n=0,\ldots,3.$ Further, 
 using the assumption (\ref{vanish}), we compute some explicit answers.
  
  
  Let us use the adjunction formula
 \be \label{A2_1}0\mapsto \cO_X(-3)\mapsto \Omega^1_Y\vert_X\mapsto \Omega^1_X\mapsto 0\ee
 as well as the short exact sequence which defines restriction of $\Omega^1_Y$ to $X$
 \be \label{A2_2}0\mapsto \Omega^1_Y\vert_X\mapsto 
 \VV_{(1,0)} \otimes \cO_{X}(-1)\oplus \VV_{(0,1)} \otimes \cO_{X}(-2)
  \mapsto \cO_X\mapsto 0\ee
  Note that (\ref{A2_2}) simply states
  $$\omega \in \Omega^1_Y\vert_X \quad \text{iff} \quad \omega=a^idt_i+b_jdU^j \quad
  \text{s. t.} \quad 2a^it_i+b_jU^j=0$$
From long exact sequences for cohomologies following from 
(\ref{A2_1},\ref{A2_2}) we find
\be \label{A2_22}I^{(1)}\Bigl(X,\cO_X(3m)\Bigr)=
 \VV_{(1,0)} \otimes I^{(0)}\bigl(X,\cO_{X}(3m-1)\bigr) +
  \VV_{(0,1)} \otimes I^{(0)}\bigl(X,\cO_{X}(3m-2)\bigr)\ee
$$- I^{(0)}\bigl(X,\cO_{X}(3m)\bigr)-I^{(0)}\bigl(X,\cO_{X}(3m-3)\bigr)$$
For example, we find {\it using the assumption} (\ref{vanish})
$$\boxed{I^{(1)}\Bigl(X,\cO_X(3)\Bigr)=2\times \VV_{(1,1)}+\VV_{(0,0)}}$$
 $$\boxed{I^{(1)}\Bigl(X,\cO_X(6)\Bigr)=2\times\Biggl(\VV_{(4,1)}+\VV_{(2,2)}+\VV_{(3,0)}+\VV_{(1,1)}\Biggr)+\VV_{(0,3)}}$$

 $$I^{(1)}\Bigl(X,\cO_X(12)\Bigr)=
2\times\Biggl(\VV_{(10,1)}+\VV_{(9,0)}+\VV_{(8,2)}+\VV_{(7,1)}+\VV_{(6,3)}+
\VV_{(5,2)}+\VV_{(4,4)}+\VV_{(3,3)}+\VV_{(2,5)}+\VV_{(1,4)}\Biggr)$$
$$+\VV_{(0,6)}$$
Independently of (\ref{A2_22}), we  computed (by listing all the sections and checking
the finiteness of their norm) the following cohomology groups
\be \label{A2_3} 
\HH^0\Bigl(X,\Omega^1_X(3)\Bigr)=2\times \VV_{(1,1)}+\VV_{(0,0)}\ee
\be \label{A2_4} \HH^0\Bigl(X,\Omega^1(6)\Bigr)=2\times\Biggl(\VV_{(4,1)}+\VV_{(2,2)}+\VV_{(3,0)}+\VV_{(1,1)}\Biggr)+\VV_{(0,3)}\ee
\be \label{A2_5}\HH^0\Bigl(X,\Omega^1_X(12)\Bigr)=
2\times\Biggl(\VV_{(10,1)}+\VV_{(9,0)}+\VV_{(8,2)}+\VV_{(7,1)}+\VV_{(6,3)}+
\VV_{(5,2)}+\VV_{(4,4)}+\VV_{(3,3)}+\VV_{(2,5)}+\VV_{(1,4)}\Biggr)\ee
$$+\VV_{(0,6)}.$$
The equations (\ref{A2_3}-\ref{A2_5}) can be put into the form
\be \label{A2_23}\HH^0\Bigl(X,\Omega^1_X\otimes \cO_X(3m)\Bigr)=
 \VV_{(1,0)} \otimes \HH^0\bigl(X,\cO_{X}(3m-1)\bigr) +
  \VV_{(0,1)} \otimes \HH^0\bigl(X,\cO_{X}(3m-2)\bigr)\ee
$$- \HH^0\bigl(X,\cO_{X}(3m)\bigr)-\HH^0\bigl(X,\cO_{X}(3m-3)\bigr)$$
Explicitly, the basis of sections in $H^0\Bigl(X,\Omega^1_X(3)\Bigr)$
is given by (in the holomorphic gauge)
\begin{enumerate}
\item $h^i_k(U^kdt_j-2t_jdU^k)$
transform in $\mathbb{V}_{(1,1)}+\VV_{(0,0)}.$
\item $C_{i[jk]}U^iU^jdU^k$  transform in $\VV_{(1,1)}.$
\end{enumerate}
Meanwhile, general $\omega \in \HH^0\Bigl(X,\Omega^{1}_X(6)\Bigr)$ is
$$\omega=a^i_{(j_1j_2j_3j_4)}\Bigl(dt_iU^{j_1}-2t_idU^{j_1}\Bigr)U^{j_2}U^{j_3}U^{j_4}+b_{(j_1j_2)}^{(ik)}t_k \Bigl(dt_iU^{j_1}-2t_i dU^{j_1}\Bigr)U^{j_2}+C_{(knm)}
dt_{i}t_{j}\epsilon^{ijk}U^nU^m$$
$$+f_{i_1i_2i_3i_4[i_5l_6]}U^{i_1}U^{i_2}U^{i_3}U^{i_4}U^{i_5}dU^{i_6}+
\tC^{pkn}\epsilon_{ijp}t_kt_nU^idU^j+\tb^{jk}_{mn}U^mU^nt_k\epsilon_{jip}U^idU^p+
d^{ijk}\,t_i t_{[j} dt_{k]}$$
where the following identification is due to $U^it_i=0$:
$$a^i_{(j_1j_2j_3j_4)}\sim a^i_{(j_1j_2j_3j_4)}+\delta^i_{(j_1}\beta_{j_2j_3j_4)},\quad
b_{(j_1j_2)}^{(ik)}\sim b_{(j_1j_2)}^{(ik)}+\gamma^{(i}_{(j_1}\delta^{k)}_{j_2)}$$
$$\tb^{jk}_{mn}\sim \tb^{jk}_{mn}+\delta^j_{(m}\alpha_{n)}^k,\quad
\tb^{jk}_{mn}\sim \tb^{jk}_{mn}+\delta^k_{(m}\alpha_{n)}^j.$$
Similarly, sections in $\HH^0(X,\Omega^1_X(12)$ are given
(in the holomorphic gauge) by

$$\omega=\epsilon^{ijp}t_idt_j\,U^{n_1}\ldots U^{n_8}\, C_{(p n_1\ldots n_8)}+
\epsilon^{ijp}t_idt_j\,t_k\,U^{n_1}\ldots U^{n_6}\, C^k_{(p n_1\ldots n_6)}+
\epsilon^{ijp}t_idt_j\,t_{k_1} t_{k_2}\,U^{n_1}\ldots U^{n_4}\, C^{k_1 k_2}_{(p n_1\ldots n_4)}$$
$$+\epsilon^{ijp}t_idt_j\,t_{k_1} t_{k_2} t_{k_3}\,U^{n_1}U^{n_2}\, C^{k_1 k_2 k_3}_{(pn_1n_2)}+\epsilon^{ijp}t_idt_j\,t_{k_1} t_{k_2} t_{k_3} t_{k_4}\, C^{k_1 k_2 k_3 k_4}_p
+$$
$$a^i_{(j_1\ldots j_{10})}\bigl(dt_iU^{j_1}-2t_idU^{j_1}\bigr)\, U^{j_2}\ldots
U^{j_{10}}+a^{(ik)}_{(j_1\ldots j_8)}\bigl(dt_iU^{j_1}-2t_idU^{j_1}\bigr)\,t_k\, U^{j_2}\ldots
U^{j_8}$$
$$+a^{(k_1k_2k_3)}_{(j_1\ldots j_6)}\bigl(dt_{k_1}U^{j_1}-2t_{k_1}dU^{j_1}\bigr)\,
t_{k_2}\,t_{k_3}\, U^{j_2}\ldots
U^{j_6}$$
$$+a^{(k_1k_2k_3k_4)}_{(j_1\ldots j_4)}\bigl(dt_{k_1}U^{j_1}-2t_{k_1}dU^{j_1}\bigr)\,
t_{k_2}\ldots t_{k_4}\, U^{j_2}\ldots
U^{j_4}+a^{(k_1\ldots k_5)}_{(j_1j_2)}\bigl(dt_{k_1}U^{j_1}-2t_{k_1}dU^{j_1}\bigr)\,
t_{k_2}\ldots t_{k_5}\, U^{j_2}$$
where coefficients are subject to identifications due to
$t_iU^i=0$ constraint.

For example:
$$C^{k_1 k_2 k_3 k_4}_p\sim C^{k_1 k_2 k_3 k_4}_p+\delta_p^{(k_1}\,B^{k_2k_3k_4)},
\quad C^{k_1 k_2 k_3}_{(pn_1n_2)}\sim C^{k_1 k_2 k_3}_{(pn_1n_2)}+
\delta_{(p}^{(k_1}\,B^{k_2k_3)}_{n_1n_2)}$$
$$ C^{k_1 k_2}_{(pn_1n_2n_3n_4)}\sim C^{k_1 k_2 }_{(pn_1n_2n_3n_4)}+
\delta_{(p}^{(k_1}\,B^{k_2)}_{n_1n_2n_3n_4)},\quad
C^k_{(p n_1\ldots n_6)}\sim C^k_{(p n_1\ldots n_6)}+\delta^k_{(p}B_{n_1\ldots n_6)}.$$

 \subsection{$L^2$ Dolbeault cohomology  of $\Omega^2_{X}(3m)$ with $m\ge 1$}
\label{appendix_three}
Here we compute $I^{(2)}\Bigl(X,\cO_X(3m)\Bigr)$ for $m\ge 1.$
We use the adjunction formula
\be \label{A3_1}
0\mapsto \cO_X(-3)\otimes \Omega^1_X\mapsto \Omega^2_Y\vert_X \mapsto \Omega^2_X\mapsto 0\ee
as well as the short exact sequence for the restriction of $\Omega^2_Y$ to $X$:
\be \label{A3_2} 0\mapsto \Omega^2_Y\vert_X \mapsto  \VV_{(0,1)}\otimes \cO_X(-2)\oplus
\VV_{(1,0)}\otimes \cO_X(-4)\oplus \Bigl (\VV_{(1,1)}+\VV_{(0,0)}\Bigr)\otimes \cO_X(-3)
\mapsto\Omega^1_Y\vert_X\mapsto 0\ee
We find
\be \label{A3_3} I^{(2)}\Bigl(X,\cO_X(3m)\Bigr)=
\VV_{(1,0)}\otimes I^{(0)}\Bigl(X,\cO_X(3m-4)\Bigr)+
\Biggl(\VV_{(1,1)}+\VV_{(0,0)}\Biggl)\otimes I^{(0)}\Bigl(X,\cO_X(3m-3)\Bigr)\ee
$$-
\VV_{(1,0)}\otimes I^{(0)}\Bigl(X,\cO_X(3m-1)\Bigr)+I^{(0)}\Bigl(X,\cO_X(3m)\Bigr)
-I^{(1)}\Bigl(X,\cO_X(3m-3)\Bigr).$$
For special cases $m=1,2$ we find, {\it using the vanishing assumption} (\ref{vanish}):
$$I^{(2)}\Bigl(X,\cO_X(3)\Bigr)=\VV_{(0,0)},\quad
I^{(2)}\Bigl(X,\cO_X(6)\Bigr)=3\VV_{(3,0)}+\VV_{(2,2)}+\VV_{(0,3)}+3\VV_{(1,1)}+\VV_{(0,0)}.$$
Independently of (\ref{A3_3}), we compute
\be \label{A5_10} \HH^0\Bigl(X,\Omega^2(3)\Bigr)=\VV_{(0,0)}\ee
\be \label{A5_11}\HH^0\Bigl(X,\Omega^2(6)\Bigr)=3\VV_{(3,0)}+\VV_{(2,2)}+\VV_{(0,3)}+3\VV_{(1,1)}+\VV_{(0,0)}.\ee
Explicitly, $\HH^0\Bigl(X,\Omega^2_X(3)\Bigr)$ is one-dimensional and generated by
$$\epsilon_{ijk}U^idU^j\wedge dU^k.$$ Meanwhile, sections in $\HH^0\Bigl(X,\Omega^2_X(6)\Bigr)$ are
$$\omega=\rho^j_i \wedge \rho^n_k \epsilon^{ikq}C_{qjn}+\rho^j_i\wedge \mu_k \tilde C_{jn}^{ki}U^n+\zeta^k\wedge \mu_n\tilde B^n_k+\alpha \zeta^k\wedge dt_k
+\mu_k\wedge dU^k\Bigl(\tilde \beta_{mnp}U^mU^nU^p+\beta_n^kU^nt_k\Bigr)$$
where we use holomorphic 1-forms well-defined on $X=T\PP^2$
$$\rho^i_j=U^jdt_i-2t_idU^j,\quad \zeta^k=\epsilon^{kij}t_idt_j,\quad \mu_k=\epsilon_{kij}U^idU^j$$
and coefficients are subject to constraints due to $t_iU^i=0.$

\subsection{$L^2$ Dolbeault cohomology  of $\Omega^3_{X}(3m)$ and $\Omega^4_{X}(3m)$ with $m\ge 1$}
\label{appendix_five}
We use  the adjunction formula
\be \label{A5_1}
0\mapsto \cO_X(-3)\otimes \Omega^2_X\mapsto \Omega^3_Y\vert_X \mapsto \Omega^3_X\mapsto 0\ee
as well as the short exact sequence for the restriction of $\Omega^3_Y$ to $X$:
\be \label{A5_2} 0\mapsto \Omega^3_Y\vert_X \mapsto
\cO_X(-3)\oplus \cO_X(-6)+\Biggl(\VV_{(0,2)}+\VV_{(1,0)}\Biggr)\otimes \cO_X(-4)
\oplus \Biggl(\VV_{(2,0)}+\VV_{(0,1)}\Biggr)\otimes \cO_X(-5)
\mapsto\Omega^2_Y\vert_X\mapsto 0\ee
Now using (\ref{A5_1}) and (\ref{A5_2}) we find
$$I^{(3)}\Bigl(\cO_X(3m+3)\Bigr)=-I^{(2)}\Bigl(\cO_X(3m)\Bigr)+
I^{(0)}\Bigl(\cO_X(3m)\Bigr)+I^{(0)}\Bigl(\cO_X(3m-3)\Bigr)-I^{(2)}\Bigl(\cO_X(3m+3)\Bigr)$$
$$+ \Biggl(\VV_{(0,2)}+\VV_{(1,0)}\Biggr)\otimes 
I^{(0)}\Bigl(\cO_X(3m-1)\Bigr)+
 \Biggl(\VV_{(2,0)}+\VV_{(0,1)}\Biggr)\otimes I^{(0)}\Bigl(\cO_X(3m-2)\Bigr)
  -I^{(1)}\Bigl(\cO_X(3m)\Bigr).$$

For example, {\it using the vanishing assumption} (\ref{vanish}), we find:
$$\boxed{I^{(3)}\Bigl(\cO_X(3)\Bigr)=-\VV_{(0,0)}}\quad \boxed{I^{(3)}\Bigl(\cO_X(6)\Bigr)=2\VV_{(1,1)}+\VV_{(0,0)}.}$$
Independently of (\ref{A5_1}) and (\ref{A5_2}), we compute
$$\HH^0\Bigl(X,\Omega^3_X(6)\Bigr)=2\VV_{(1,1)}+\VV_{(0,0)}$$
where general section of $\HH^0\Bigl(X,\Omega^3_X(6)\Bigr)$ is given by
$$\omega=\Bigl(\mu_k\wedge d\zeta^p+2d(\mu_k)\wedge \zeta^p\Bigr)C^k_p+
B_i^k \rho^i_k \wedge d(\mu_m)U^m$$
 with $B^i_k\sim B^i_k +\delta^i_k \gamma.$
 
Similarly, for $\Omega^4_X$ we use the adjunction formula
\be \label{A6_1}
0\mapsto \cO_X(-3)\otimes \Omega^3_X\mapsto \Omega^4_Y\vert_X \mapsto \Omega^4_X\mapsto 0\ee
as well as the short exact sequence for the restriction of $\Omega^4_Y$ to $X$:
\be \label{A6_2} 0\mapsto \Omega^4_Y\vert_X \mapsto \VV_{(0,1)}\otimes \cO_X(-5)\oplus
 \Biggl(\VV_{(0,0)}+\VV_{(1,1)}\Biggr)\otimes \cO_X(-6)\oplus
\VV_{(1,0)}\otimes \cO_X(-7)
\mapsto\Omega^3_Y\vert_X\mapsto 0\ee
This allows to identify
\be \label{A6_12}\Omega^4_X=\cO_X(-9)\ee
and compute
$$I^{(4)}\Bigl(\cO_X(3m)\Bigr)=I^{(0)}\Bigl(X,\cO_X(3m-9)\Bigr).$$
The {\it vanishing assumption} (\ref{vanish}) implies \footnote{In the first equation we used Kodaira-Serre duality for $L^2$ Dobeault cohomology.}

$$\boxed{I^{(4)}\Bigl(X,\cO_X(3)\Bigr)=I^{(0)}\Bigl(X,\cO_X(-3)\Bigr)=0},\quad \boxed{I^{(4)}\Bigl(X,\cO_X(6)\Bigr)=I^{(0)}\Bigl(X,\cO_X(-3)\Bigr)=0}$$

Collecting the pieces together we find the Euler characters of $\cO_X(3)$ and $\cO_X(6)$ bundles
$${\bf I}_{\cN=4}\Bigl(X,\cO_X(3)\Bigr)=\VV_{(3,0)}-\VV_{(1,1)}+\VV_{(0,0)}$$
$${\bf I}_{\cN=4}\Bigl(X,\cO_X(6)\Bigr)=\VV_{(6,0)}-\VV_{(4,1)}+
\VV_{(0,3)}+\VV_{(3,0)}-\VV_{(1,1)}.$$

\section{Some evidence in support of the `vanishing assumption'}
Here we explain how to compute $\HH^i\Bigr(X,\cO_X(n)\Bigl)$
for $i>0$ and give some explicit examples in support of the vanishing assumption 
(\ref{vanish}).
\subsection{$\HH^1\Bigl(X,\cO_X(n)\Bigr)$}
Let us define 
$$\ts={\tx \over Y^2}, \quad \tx=t_a \bar t^a, \quad Y={\overline U}_a U^a$$
so that  in $U^1\ne 0$ patch with inhomogenous coordinates $z^1={U^2\over U^1}$
and $z^2={U^3\over U^1}$
on $\PP^2$ we solve $t_1=-(t_2 z^1+t_3 z^2)$ and write
$$\ts ={\tx\over y^2},\quad \tx=t_2 {\overline \alpha}^2+ t_3 {\overline \alpha}^3,\quad
y=1+\vert z_1\vert^2+\vert z_2\vert^2$$
with
 $${\overline \alpha}^2=\bar t^2 +z^1 
 (\bar t^2 \bar z_1+\bar t^3 \bar z_2)$$
 $${\overline \alpha}^3=\bar t^3 +z^2 
 (\bar t^2 \bar z_1+\bar t^3 \bar z_2).$$

Let us write general $\omega \in \Omega^{(0,1)}_X\otimes \cO_X(n)$ as
$$\omega=\sum_{I=1}^4\alpha_I {\bar e}_I$$
where 
$$e_1=\partial \ts,\quad e_2={t_adU^a\over Y^{3/2}}$$
 $$e_3={\epsilon^{acd}{\overline U}_at_c dt_d\over  Y^{5/2}},\quad
 e_4=
 {\epsilon_{abc}\bar t^a U^b dU^c\over y^2}$$In $U^1\ne 0$ patch we find
$$e_1={\partial \tx \over y^2}-{2\partial y\over y}\ts,\quad e_2={t_2dz^1+t_3dz^2\over y^{3/2}}
 $$
 $$e_3={t_2dt_3-t_3dt_2\over y^{3/2}}+{t_2\bar z_2-t_3\bar z_1\over y}e_2,\quad
 e_4={\bar \alpha^3 dz^1-
 \bar \alpha^2 dz^2\over y^2}.$$

Note that ${U^a\over \sqrt{y}}$ and ${t_a\over y}$ are multiplied by
 phase under the  $C^*$ action on $\mathbb{P}^2_{\vec U}$, so 
 $$e_1 \in \Gamma\Bigl(\Omega_X^{1,0}\Bigr),\quad 
 e_2 \in \Gamma\Bigl(\Omega_X^{1,0}(3)\Bigr),\quad 
 e_3 \in \Gamma\Bigl(\Omega_X^{1,0}(3)\Bigr),\quad 
 e_4 \in \Gamma\Bigl(\Omega_X^{1,0}\Bigr).$$ 
For this reason
$$\alpha_{1,4}\in \Gamma\Bigl(\cO_X(n)\Bigr),\quad 
\alpha_{2,3}\in \Gamma\Bigl(\cO_X(n+3)\Bigr).$$
Let us consider differential equations
$$\overline{D}^{(n)}\omega=0,\quad D^{(n)}(*\omega)=0$$ with
\be\label{con_i} D^{(n)}=\nabla + {h^{'(n)}\over h^{(n)}}e_1,\quad 
\overline{D}^{(n)}=\overline{\nabla}-{h^{'(n)}\over h^{(n)}}\bar e_1\ee
Here 
$$\nabla=\partial -{n\over 2}{\partial y\over y},\quad
\overline{\nabla}=\delbar +{n\over 2}{\delbar y\over y}$$
and
 \be \label{con_ii}h^{(n)}(\ts)\sim \ts^{-{n\over 4}} \,\,\,\, \ts \mapsto \infty,\quad h^{(n)}\sim 1\,\,\,\,
 \ts \mapsto 0\ee
  
 We define $\beta_{I;K}$ and $\gamma_{I;K}$ as
$$ \overline{\nabla}\alpha_I=\sum_{K}\beta_{I;K}\overline{e}_K$$
$$ \nabla\alpha_I=\sum_{K}\gamma_{I;K} e_K$$
and use
 $$\nabla e_1=0, \quad \nabla \overline{e}_2=0, \quad
 \nabla \overline{e}_1={1\over \ts}
 e_1 \wedge
  \overline{e}_1- e_2 \wedge \overline{e}_2+
  {1\over \ts}e_3 \wedge 
  \overline{e}_3-2e_4 \wedge \overline{e}_4$$
  $$\nabla e_4=0,\quad  \nabla \overline{e}_4={1\over \ts}e_1\wedge \overline{e}_4+
  {1\over \ts}e_3\wedge\overline{e}_2,
  \quad \nabla e_2={1\over \ts}e_1\wedge e_2 -{1\over \ts}e_3\wedge e_4$$
  $$\nabla e_3={2\over \ts}e_1\wedge e_3, \quad
 \nabla \overline{e}_3=-e_4 \wedge \overline{e}_2$$
  to show that $\overline{D}^{(n)}\omega=0$ is equivalent to
  \be \label{harm1}\beta_{1;4}=\beta_{4;1}-{h^{'(n)}\over h^{(n)}}\alpha_4,\quad \beta_{2;3}=\beta_{3;2},\quad 
  \beta_{2;4}=\beta_{4;2}\ee
  $$\alpha_2=\ts\Bigl(\beta_{1;2}-\beta_{2;1}+{h^{'(n)}\over h^{(n)}}\alpha_2\Bigr),\quad
 2\alpha_3=\ts\Bigl(\beta_{1;3}-\beta_{3;1}+{h^{'(n)}\over h^{(n)}}\alpha_3\Bigr),\quad
 \alpha_2=\ts\bigl(\beta_{4;3}-\beta_{3;4}\bigr)$$
  while $D^{(n)}(*\omega)=0$ is equivalent to
  \be \label{harm2}\alpha_1\Biggl(\Bigl({5 \over \ts}+ {h^{'(n)}\over h^{(n)}}\Bigr)\tf_2\tf_3\tf_4+\bigl(\tf_2\tf_3\tf_4\bigr)'\Biggr)+
  \sum_{K=1}^4\gamma_{K;K} \prod_{J\ne K}\tf_J=0\ee
 Here functions $\tf_i(\ts)$ are coefficients of expanding K\"ahler form (\ref{k_form})
 on $X.$
Finally, we have to compute the norm squared
\be \label{norm}\vert \vert \omega \vert\vert^2=\int_X \omega \wedge *\overline{\omega}=
\sum_{K=1}^4\int_X {\vert \alpha_K\vert^2 \, vol_X\over \tf_K} \ee
where the volume form can be written as\footnote{$t_2=t_3v,\,t_3=r_3e^{i\varphi},\,\tx=r_3^2\bigl(1+\vert v\vert^2+\vert z^1 v+z^2\vert^2\bigr)$}
$$vol_X=\ts^5 \prod_{j=1}^4 \tf_j(\ts)\,d\ts\wedge d\varphi \wedge {dv \wedge d \bar v\over \bigl(1+\vert v\vert^2+\vert z^1 v+z^2\vert^2\bigr)^2}
\wedge {dz^1\wedge d\bar z_1\wedge dz^2\wedge d\bar z_2\over y^2}$$

Note that $\alpha_K$ can be written as polynomials in
\be \label{def_nu}\nu^i_1={U^i\over y^{1/2}},\quad \nu^i_2={\bar t^i\over y},\quad \nu^i_3={\epsilon^{ijk}\overline{U}_j t_k\over y^{3/2}}\ee
and their conjugates. In computing $\beta_{I:K}$ and $\gamma_{I;K}$ we
use
 \be \label{b_2} \ts \, \nabla \Bigl(\nu_1^i\Bigr)=\nu_2^i e_2 -
  \nu_3^i e_4,\quad \overline{\nabla}\Bigl(\nu_1^i\Bigr)=0,\quad
 \nabla\Bigl(\nu_2^i\Bigr)=0,\quad \ts \, \overline{\nabla} \Bigl(\nu^i_2\Bigr)=\nu_2^i \bar e_1 - \nu_1^i \,\ts\, \bar e_2+
  \nu_3^i \bar e_3 \ee
  \be \label{b_3} \ts \nabla\Bigl(\nu_3^i\Bigr)=\nu_3^i e_1-\nu_2^i e_3,\quad
  \ts \overline{\nabla}\Bigl(\nu_3^i\Bigr)=\nu_1^i \bar e_4\ee
  
\subsection{$\HH^1\Bigl(X,\cO_X(1)\Bigr)$}
 Let us first look for $\omega \in\HH^1\Bigl(X,\cO_X(1)\Bigr) $ that transforms
 in irreducible representation $\VV_{(1,0)}$. This implies $\alpha_2^i=\alpha^i_3=0$ and
 $$\alpha_1^i=c_1(\ts)\nu_1^i +b_1(\ts)\nu_3^i,\quad
  \alpha_4^i=c_4(\ts)\nu_1^i +b_4(\ts)\nu_3^i$$
  
  Now we use (\ref{b_2}) and (\ref{b_3})
  to compute $\beta^i_{1;2}=\beta^i_{1;3}=\beta^i_{4;2}=\beta^i_{4;3}=0$ and
$$\beta^i_{1;4}=b_1(\ts) \nu^i_1,\quad \beta^i_{4;1}=c_4' \, \nu^i_1+b_4' \, \nu^i_3$$
as well as
$$\gamma^i_{1;1}=c_1' \, \nu^i_1+
b_1' \,\nu^i_3,\quad \gamma^i_{4;4}=-{c_4\over \ts}\nu_3^i$$
Now from (\ref{harm1}) and (\ref{harm2}) we obtain
$$b_1=c_4'-{h^{(1)'}\over h^{(1)}}c_4, \quad b_4'-{h^{(1)'}\over h^{(1)}}b_4=0$$
$$c_1\Biggl({5 \tf_2 \tf_3 \tf_4\over \ts}+ \Bigl(\tf_2 \tf_3 \tf_4\Bigr)'\Biggr)+
\Bigl(c_1'+{h^{(1)'}\over h^{(1)}}c_1\Bigr)\tf_2 \tf_3 \tf_4=0$$
$$b_1\Biggl({5 \tf_2 \tf_3 \tf_4\over \ts}+ \Bigl(\tf_2 \tf_3 \tf_4\Bigr)'\Biggr)+
\Bigl(b_1'+{h^{(1)'}\over h^{(1)}}b_1\Bigr)\tf_2 \tf_3 \tf_4-{c_4 \tf_1 \tf_2\tf_3 \over \ts}=0$$
In the limit $\ts \mapsto \infty $ (i.e. $s\mapsto 0$) both $b_4$ and $c_1$ behave as
$\ts^{-1/4}$ while the two solutions for $c_4$ behave as $\ts^{-1/4}$ and $\ts^{3/4}$
in this limit. We checked using the norm (\ref{norm}) that all of these solutions have divergent
contribution to their norm squared from the region $\ts \mapsto \infty.$
Hence, we conclude that there is no representation $\VV_{(1,0)}$ in $\HH^1\Bigl(X,\cO_X(1)\Bigr).$

 Now let us  look for $\omega \in\HH^1\Bigl(X,\cO_X(1)\Bigr) $ that transforms
 in irreducible representation $\VV_{(0,2)}$:
 $$\alpha_J=a_J(\ts) \bar \nu_{1\, (j}\bar \nu_{2\, k)}+b_J(\ts) \bar \nu_{2\, (j}\bar \nu_{3\, k)}\quad J=1,4;
 \quad \alpha_J=a_J(\ts)  \bar \nu_{2\, j}\bar \nu_{2\, k}\quad J=2,3.$$
Then, we compute $\beta_{I:K}$ and $\gamma_{K;K}$(we only write non-zero components)
$$\beta_{J;1}=a'_J \, \bar \nu_{1\, (j}\bar \nu_{2\, k)}+
\Bigl(b'_J+{b_J\over \ts}\Bigr)\bar \nu_{2\, (j}\bar \nu_{3\, k)}\quad J=1,4; \quad
\beta_{J;1}=a'_J \, \bar \nu_{2\, j}\bar \nu_{2\, k} \quad J=2,3;$$
$$\beta_{J;2}={a_J\over \ts}\, \bar \nu_{2\, j}\bar \nu_{2\, k},\quad
 \beta_{J;3}=-{b_J\over \ts}\, \bar \nu_{2\, j}\bar \nu_{2\, k},\quad \beta_{J;4}=-{a_J\over \ts} \bar \nu_{2\, (j}\bar \nu_{3\, k)}\quad J=1,4;$$
$$\gamma_{1;1}=\Bigl(a'_1+{a_1\over \ts}\Bigr)\, \bar \nu_{1\, (j}\bar \nu_{2\, k)}+
\Bigl(b'_1+{b_1\over \ts}\Bigr)\, \bar \nu_{2\, (j}\bar \nu_{3\, k)},\,\, \gamma_{2;2}=-2a_2  \bar \nu_{1\, (j}\bar \nu_{2\, k)},\,\,  \gamma_{3;3}={2a_3\over \ts}  \bar \nu_{2\, (j}\bar \nu_{3\, k)}, \,\,
\gamma_{4;4}={b_4\over \ts}\,\bar \nu_{1\, (j}\bar \nu_{2\, k)}$$
Now from (\ref{harm1}) and (\ref{harm2}) we obtain
$$a_4=0,\quad a_2=-b_4,\quad a_1=a_2+\ts \Bigl(a'_2-{h^{(1)'}\over h^{(1)}}a_2\Bigr),\quad
b_1=-2a_3-\ts \Bigl(a'_3-{h^{(1)'}\over h^{(1)}}a_3\Bigr)$$
$$b_1\Biggl( \Bigl({6\over \ts}+{h^{(1)'}\over h^{(1)}}\Bigr) \tf_2\tf_3\tf_4+\Bigl(\tf_2\tf_3\tf_4\Bigr)'\Biggr)+b'_1 \tf_2\tf_3\tf_4+{2a_3\over \ts}\tf_1\tf_2\tf_4=0$$
$$a_1\Biggl( \Bigl({6\over \ts}+{h^{(1)'}\over h^{(1)}}\Bigr) \tf_2\tf_3\tf_4+\Bigl(\tf_2\tf_3\tf_4\Bigr)'\Biggr)+a'_1 \tf_2\tf_3\tf_4-2a_2\tf_1\tf_3\tf_4+{b_4\over \ts} \tf_1\tf_2\tf_3=0.$$
We checked that there are no well-behaved solutions with finite norm.
\subsection{$\HH^1\Bigl(X,\cO_X(-1)\Bigr)$}
Let us look for $\omega \in \HH^1\Bigl(X,\cO_X(-1)\Bigr)$ that transforms in irreducible representation
$\VV_{(2,0)}$:
$$\omega=\Bigl(K(\ts) \nu_1\nu_2+L(\ts)\nu_2\nu_3\Bigr) \bar e_1+\Bigl(A(\ts) \nu_1\nu_3+B(\ts)\nu_1^2+E(\ts)\nu_3^2\Bigr) \bar e_2+ 
\Bigl(\tilde A(\ts) \nu_1\nu_3+\tilde B(\ts)\nu_1^2+\tilde E(\ts)\nu_3^2\Bigr) \bar e_3$$
$$+
\Bigl(\tilde K(\ts) \nu_1\nu_2+\tilde L(\ts)\nu_2\nu_3\Bigr) \bar e_4$$
where $\nu_1,\nu_2,\nu_3$ are defined in (\ref{def_nu}) and
indices are suppressed. For example, 
$$ K(\ts)\nu_1 \nu_2 =K_{ij}\nu_1^i\nu_2^j \quad K_{ij}=K_{ji}.$$
We compute
$$\beta_{2;1}=A' \nu_1\nu_3+B'\nu_1^2+E'\nu_3^2,\quad
\beta_{3;1}=\tilde A' \nu_1 \nu_3+\tilde B' \nu_1^2+\tilde E'\nu_3^2$$
$$\beta_{4:1}=(\tilde K'+\ts^{-1}\tilde K)\nu_1\nu_2+
(\tilde L'+\ts^{-1}\tilde L)\nu_2\nu_3$$
$$\beta_{1;2}=-(K\nu_1^2+L\nu_1 \nu_3),\quad \beta_{1;3}=\ts^{-1}*(K\nu_1\nu_3+L\nu_3^2),\quad \beta_{1;4}=\ts^{-1}L\nu_1\nu_2$$
$$\beta_{4;2}=-(\tilde K\nu_1^2+\tilde L\nu_1 \nu_3),\quad 
\beta_{2,4}=\ts^{-1}*(A\nu_1^2+2E\nu_1\nu_3)$$
$$\beta_{4;3}=\ts^{-1}*(\tilde K\nu_1\nu_3+\tilde L\nu_3^2),\quad 
\beta_{3,4}=\ts^{-1}*(\tilde A\nu_1^2+2\tilde E\nu_1\nu_3).$$
Plugging this into (\ref{harm1}) gives rise to
$$A=E=L=\tilde L=\tilde K=\tilde E=0,\quad B=-\tilde A$$
$$K=-\ts \nabla_{\ts}^- \tilde A +2 \tilde A,\quad
K=-\nabla_{\ts}^- \tilde A +{\tilde A\over \ts}
$$
$$2\tilde B+\ts \nabla_{\ts}^- \tilde B=0.$$
Note that the two different expression for $K$ imply the
first order differential equation for $\tilde A$. Further, we checked that the solution for $\tilde B$ has divergent norm.
Meanwhile,
$$\gamma_{1;1}=K' \nu_1\nu_2,\quad \gamma_{2;2}=-2{\tilde A \over \ts}\nu_1\nu_2,\quad \gamma_{3;3}=-{\tilde A \over \ts}\nu_1\nu_2$$
and (\ref{harm2}) gives a second order equation for $\tilde A.$
The only solution of the two differential equations is $\tilde A=0.$
We conclude that there is no $(2,0)$ in $\HH^1\bigl(X,\cO_X(-1)\bigr).$ 
\subsection{$\HH^2\Bigl(X,\cO_X(-1)\Bigr)$}
Let us look for representation $\VV_{(0,1)}$ in $\HH^2\Bigl(X,\cO_X(-1)\Bigr)$:
$$\omega=\bigl(K(\ts) \bar \nu_1 +L(\ts) \bar \nu_3\bigr)\,\bar e_1 \wedge \bar e_4+ 
N(\ts) \bar \nu_2\, \bar e_1 \wedge \bar e_2+P(\ts)\bar \nu_2 \,\bar e_1 \wedge \bar e_3+
G(\ts) \bar \nu_2\, \bar e_2 \wedge \bar e_4+M(\ts) \bar \nu_2 \,\bar e_3 \wedge \bar e_4$$
So that
$$*\omega=\Bigl(\hat K(\ts) \bar \nu_1 + \hat L(\ts) \bar \nu_3 \Bigr)\, \bar e_1 \wedge \bar e_4\wedge e_2\wedge \bar e_2\wedge e_3 \wedge \bar e_3
+ \hat N(\ts) \bar \nu_2 \, \bar e_1 \wedge \bar e_2\wedge e_3\wedge \bar e_3\wedge e_4 \wedge \bar e_4$$
$$+ \hat P(\ts) \bar \nu_2 \, \bar e_1 \wedge \bar e_3
\wedge e_2\wedge \bar e_2\wedge e_4 \wedge \bar e_4$$
$$+\hat G(\ts) \bar \nu_2\,  \bar e_2 \wedge \bar e_4
\wedge e_1\wedge \bar e_1\wedge e_3 \wedge \bar e_3
+ \hat M(\ts) \bar \nu_2 \, \bar e_3 \wedge \bar e_4\wedge e_1\wedge \bar e_1\wedge e_2 \wedge \bar e_2$$
where
$$\hat K=K \tf_2 \tf_3,\quad \hat N=N \tf_3 \tf_4,\quad \hat L=L \tf_2\tf_3,\quad
\hat P=P \tf_2\tf_4,\quad \hat G=G\tf_1\tf_3, \quad \hat M=M\tf_1\tf_2.$$
 Imposing $\overline{D}^{(-1)}\omega=0,\quad D^{(-1)}\bigl(*\omega\bigr)=0$ gives
\be \label{dbar} \ts \nabla_{\ts}^- G +G=K,\quad 
\ts \hat G=\ts \nabla_{\ts}^+ \hat K +5\hat K,  \quad L=N=M=P=0\ee
We checked that there no solutions for $G,K$ with finite norm. We conclude
that there is no $\VV_{(0,1)}$ in $\HH^2\Bigl(X,\cO_X(-1)\Bigr).$

\subsection{$\HH^2\Bigl(X,\cO_X(-2)\Bigr)$}
Let us look for representation $\VV_{(1,0)}$ in $\HH^2\Bigl(X,\cO_X(-2)\Bigr)$:
$$\omega=A(\ts) \nu_2 \bar e_1 \wedge \bar e_4+ (B(\ts) \nu_1+C(\ts)\nu_3) \bar e_1 \wedge \bar e_2+(D(\ts) \nu_1+E(\ts)\nu_3) \bar e_1 \wedge \bar e_3+$$
$$(K(\ts) \nu_1+L(\ts)\nu_3) \bar e_2 \wedge \bar e_4+(M(\ts) \nu_1+N(\ts)\nu_3) \bar e_3 \wedge \bar e_4$$
So that
$$*\omega=\hat A(\ts) \nu_2 \bar e_1 \wedge \bar e_4\wedge e_2\wedge \bar e_2\wedge e_3 \wedge \bar e_3+
+ (\hat B(\ts) \nu_1+\hat C(\ts)\nu_3) \bar e_1 \wedge \bar e_2\wedge e_3\wedge \bar e_3\wedge e_4 \wedge \bar e_4$$
$$+(\hat D(\ts) \nu_1+\hat E(\ts)\nu_3) \bar e_1 \wedge \bar e_3
\wedge e_2\wedge \bar e_2\wedge e_4 \wedge \bar e_4+$$
$$(\hat K(\ts) \nu_1+\hat L(\ts)\nu_3) \bar e_2 \wedge \bar e_4
\wedge e_1\wedge \bar e_1\wedge e_3 \wedge \bar e_3+(\hat M(\ts) \nu_1+\hat N(\ts)\nu_3) \bar e_3 \wedge \bar e_4\wedge e_1\wedge \bar e_1\wedge e_2 \wedge \bar e_2$$
where
$$\hat A=A \tf_2 \tf_3,\quad \hat B=B \tf_3 \tf_4,\quad \hat C=C\tf_3\tf_4,\quad
\hat D=D \tf_2\tf_4,\quad \hat E=E\tf_2\tf_4,$$
$$ \hat K=K\tf_1\tf_3,\quad
 \hat L=L\tf_1\tf_3,\quad \hat M=M\tf_1\tf_2,\quad \hat N=N\tf_1\tf_2.$$
 Imposing $\overline{D}^{(-2)}\omega=0,\quad D^{(-2)}*\omega=0$ gives
 $$B=M=E=0$$
 \be \label{type_i} \ts \nabla_{\ts}^{-}L+L=0,\quad \ts \nabla_{\ts}^{+}\hat D+3\hat D=0\ee
 \be \label{type_ii} \ts \nabla_{\ts}^{+}\hat A+5\hat A=\hat N-\hat K,\quad
 \ts \nabla_{\ts}^{+}\hat C+5\hat C=-\hat N-\hat K \ee
 $$(1+\ts)A=\ts \nabla_{\ts}^{-}(N-K)+2N-K,\quad
 (1+\ts)C=-\ts \nabla_{\ts}^{-}K-K-\ts \Bigl(\ts \nabla_{\ts}^{-}N+2N\Bigr)$$
 We checked that solutions for $L$ and $D$ have divergent norm. Moreover,
 the system of equations (\ref{type_ii}) has only one dimensional family of
 solutions\footnote{The other solutions at $\ts \mapsto 0$ are either too divergent
 $K\sim \ts^{-4}$ or non-analytic in $\ts$.} with good properties
 at $\ts \mapsto 0$:
 $$N=A={C_0\over \ts},\quad K=-{4C_0\over \ts},\quad C=-C_0$$
 However, we checked using Mathematica, that this solution interpolates into
 a solution with $C \sim \ts^{-5/4}$ at $\ts \mapsto \infty$ which has divergent norm.
 We conclude that there is no $\VV_{(1,0)}$ in $\HH^2\Bigl(X,\cO_X(-2)\Bigr).$

 \subsection{$\HH^j\Bigl(X,\cO_X(3)\Bigr)$ for $j=2,3,4$}
 Let us first look for $\omega \in \HH^2\Bigl(X,\cO_X(3)\Bigr)$ that transforms in representation $\VV_{(3,0)}.$ General ansatz is given by
$$\omega=\Biggl(C(\ts)\nu_1^3+B(\ts)\nu_1^2\nu_3+E(\ts)\nu_1\nu_3^2+
\nu_3^3F(\ts)\Biggr)\bar e_1 \wedge \bar e_4$$
$\overline{D}^{(3)}\omega=0$ is satisfied automatically, but $D^{(3)}(*\omega)=0$
gives
$$C=B=E=0,\quad \ts \nabla_{\ts}^+\hat F + 8\hat F=0$$
where
$\hat F=F\tf_2\tf_3.$ 
We checked that the solution for $F$ has divergent norm.

Now we look for $\VV_{(1,1)}$ in $\HH^2\Bigl(X,\cO_X(3)\Bigr).$ General ansatz is given by
$$\omega=a(\ts)\nu_1 \bar \nu_2 \bar e_1 \bar e_4$$
$\overline{D}^{(3)}\omega=0$ is satisfied automatically, but $D^{(3)}(*\omega)=0$
gives
$$\ts \nabla_{\ts}^+\hat a + 6\hat a=0\quad \hat a=a \tf_2\tf_3$$
We checked that the solution for $a$ has divergent norm.
We conclude that there is neither $\VV_{(3,0)}$ nor
$\VV_{(1,1)}$ in $\HH^2\Bigl(X,\cO_X(3)\Bigr).$
Similarly, there is no $\VV_{(0,0)}$ in $\HH^2\Bigl(X,\cO_X(3)\Bigr).$

Moreover, there are no $\VV_{(3,0)},\VV_{(1,1)},\VV_{(0,0)}$ in $\HH^j\Bigl(X,\cO_X(3)\Bigr)$ for $j=3,4$ since one cannot even write down an ansatz for $\omega$
in these cases.

\subsection{$\HH^2\Bigl(X,\cO_X(-3)\Bigr)$}
Let us look for $\omega \in \HH^2\Bigl(X,\cO_X(-3)\Bigr)$ that transforms in irreducible representation
$\VV_{(0,0)}$:
$$\omega=a(\ts) \bar e_1 \wedge \bar e_2+ b(\ts) \bar e_1 \wedge \bar e_3+
c(\ts) \bar e_2\wedge \bar e_4+ d(\ts) \bar e_3 \wedge \bar e_4$$
Then, $\overline{D}\omega=0$ gives
$$\ts \Bigl(c'-{h^{(-3)'}\over h^{(-3)}}c\Bigr)+c=0,\quad
\ts \Bigl(d'-{h^{(-3)'}\over h^{(-3)}}d\Bigr)+2d+a=0$$
We further use
$$*\Bigl(\bar e_I\wedge \bar e_J\Bigr)=\bar e_I \wedge \bar e_J \wedge \prod_{K\ne I, \, K\ne J}\tf_K(\ts)
e_K\wedge \bar e_K$$
to show that $D(*\omega)=0$ is equivalent to
$$ \Bigl(a \tf_3\tf_4\Bigr)'+\Biggl({4\over \ts}+{h^{(-3)'}\over h^{(-3)}}\Biggr)a \tf_3\tf_4+{d\over \ts}\tf_1\tf_2=0$$
$$ \Bigl(b \tf_2\tf_4\Bigr)'+\Biggl({3\over \ts}+{h^{(-3)'}\over h^{(-3)}}\Biggr)b \tf_2\tf_4=0.$$
In the limit $\ts \mapsto \infty $ (i.e. $s\mapsto 0$) 
$$c \sim \ts^{-1/4}, \quad b \sim \ts^{-5/4}\quad \ts \mapsto \infty$$
We checked using the norm (\ref{norm}) that both $b$ and $c$ solutions have divergent
contribution to their norm squared from the region $\ts \mapsto \infty.$
The two solutions for $a$ behave as
$$a\sim C_1\ts^{-7/4}+C_2\ts^{-3/4}$$ 
The solution with $C_2\ne 0$ has divergent contribution to the norm squared
from $\ts \mapsto \infty$. While in the other limit
$$a \sim C_3 \ts^{-3}+C_4\ts^{-1} \quad \ts \mapsto 0$$
and solution with  $C_3\ne 0$ has divergent contribution to the norm squared
from $\ts \mapsto 0$. We further checked, using Mathematica, that
a good solution at $\ts \mapsto 0$ behaves badly at $\ts \mapsto \infty.$
Hence, we conclude that there is no representation $\VV_{(0,0)}$ in $\HH^2\Bigl(X,\cO_X(-3)\Bigr).$

\section{Useful formulae for computing the norm}
We work in a patch $U^1\ne 0$ and introduce 'polar coordinates':
$$t_2=vt_3,\quad t_3=\vert t_3 \vert e^{i\varphi},\quad z_1=r_1e^{i\phi_1},\quad z_2=r_2e^{i\phi_2},\quad v=r_ve^{i\phi_v}$$
and denote
$$a=1+T_2+T_v(1+T_1),\quad b=2r_1r_2r_v ,\quad
\Phi=\phi_2-\phi_1-\phi_v,\quad y=1+T_1+T_2$$
where
$$T_v=r_v^2,\quad T_1=r_1^2,\quad T_2=r_2^2.$$
To evaluate the integrals arising in the computation of the norm we use
$$\int_{0}^{2\pi}{d\Phi\over a+b cos \Phi}={2\pi  \over (a^2-b^2)^{1/2}},\quad \int_{0}^{2\pi}{d\Phi\over \bigl(a+b cos \Phi\bigr)^2}={2\pi a \over (a^2-b^2)^{3/2}},\quad a>b$$
$$\int_{0}^{2\pi}{d\Phi\over \bigl(a+b cos \Phi\bigr)^3}={\pi (b^2+2a^2) \over (a^2-b^2)^{5/2}},\quad 
\int_{0}^{2\pi}{d\Phi\over \bigl(a+b cos \Phi\bigr)^4}={\pi (2a^3+3ab) \over (a^2-b^2)^{7/2}},\quad a>b$$
$$\int_{0}^{\infty}{T_v dT_v \over \bigl(\beta T_v^2+2\gamma T_v+\delta\bigr)^{3/2}}=\half {1\over y(1+T_1)},\quad
\int_{0}^{\infty}{dT_v \over \bigl(\beta T_v^2+2\gamma T_v+\delta\bigr)^{3/2}}=\half {1\over y(1+T_2)}$$
$$\int_{0}^{\infty}{T^2_v dT_v \over \bigl(\beta T_v^2+2\gamma T_v+\delta\bigr)^{5/2}}={1\over 12y^2(1+T_1)},\quad
\int_{0}^{\infty}{T_vdT_v \over \bigl(\beta T_v^2+2\gamma T_v+\delta\bigr)^{5/2}}= {1\over 12y^2(1+T_2)}$$
where
$$\beta=(1+T_1)^2,\quad \gamma=1+T_1+T_2-T_1T_2,\quad \delta=(1+T_2)^2.$$

\end{document}